\documentclass[copyright,creativecommons]{eptcs}
\providecommand{\event}{QPL 2026}

\usepackage{amsmath,amssymb,amsthm}
\usepackage{mathrsfs}  
\usepackage{stmaryrd}
\let\parr\bindnasrepma
\usepackage{tikz}
\usepackage{tikz-cd}
\usetikzlibrary{positioning, calc}
\usepackage{enumitem}
\setlist{nosep,leftmargin=*}
\usepackage{mathtools}
\usepackage{comment}

\newif\ifconfversion
\confversiontrue

\ifconfversion
  \AtBeginDocument{%
    \setlength{\abovedisplayskip}{2pt plus 1pt minus 1pt}%
    \setlength{\belowdisplayskip}{2pt plus 1pt minus 1pt}%
    \setlength{\abovedisplayshortskip}{0pt plus 1pt minus 1pt}%
    \setlength{\belowdisplayshortskip}{0pt plus 1pt minus 1pt}%
    \setlength{\topsep}{2pt}%
    \setlength{\partopsep}{0pt}%
  }
  \makeatletter
  \def\thm@space@setup{%
    \thm@preskip=2pt plus 1pt minus 1pt
    \thm@postskip=\thm@preskip}
  \makeatother
\fi

\newif\ifshowproofs
\showproofstrue

\ifconfversion
  \showproofsfalse
\fi

\ifshowproofs\else
  \excludecomment{proof}
\fi

\newcommand{\extendedonly}[1]{\ifconfversion\else#1\fi}

\usepackage{iftex}
\ifpdf
  \usepackage{underscore}
  \usepackage[T1]{fontenc}
\else
  \usepackage{breakurl}
\fi

\newtheorem{definition}{Definition}[section]
\newtheorem{theorem}[definition]{Theorem}
\newtheorem{lemma}[definition]{Lemma}
\newtheorem{proposition}[definition]{Proposition}
\newtheorem{corollary}[definition]{Corollary}

\theoremstyle{definition}

\numberwithin{equation}{section}

\newcommand{\Perm}{\mathrm{Perm}}

\newcommand{\QC}{\mathbb{QC}}

\newcommand{\id}{\mathrm{id}}

\DeclareMathOperator{\Pos}{Pos}
\DeclareMathOperator{\Neg}{Neg}

\DeclareMathOperator{\ev}{ev}

\newcommand{\CC}{\mathbb{C}}
\newcommand{\NN}{\mathbb{N}}
\newcommand{\RR}{\mathbb{R}}

\newcommand{\ii}{\mathrm{i}}
\newcommand{\ket}[1]{\lvert #1\rangle}
\newcommand{\bra}[1]{\langle #1\rvert}
\newcommand{\Unit}{\mathbf{1}}

\title{Essential Unitarity for Higher-Order Quantum Computation\\
(EXTENDED ABSTRACT)}
\author{
Samson Abramsky
\institute{University College London\\
London, UK}
\email{s.abramsky@ucl.ac.uk}
\and
Radha Jagadeesan
\institute{DePaul University\\
Chicago, USA}
\email{rjagadee@depaul.edu}
}
\def\titlerunning{Essential Unitarity for Higher-Order Quantum Computation}
\def\authorrunning{S.~Abramsky \& R.~Jagadeesan}

\begin{document}
\maketitle

\begin{abstract}
We develop a boundary-centric
semantic framework for higher-order quantum computation, building on
the Kelly--Laplaza description of compact closure and Abramsky's
execution account. In the semantic carrier \(\Perm(\CC)\), morphisms
are complex-linear combinations of polarized boundary linkings, composed
by execution. Finite-family addresses provide coherent control over
finite-level quantum registers (qudits) while retaining the multiplicative
boundary structure.

We identify \emph{essential unitarity}, a boundary condition extending
ordinary unitarity to higher-order interfaces. On positive qudit
registers it coincides with ordinary matrix unitarity; at higher order
it expresses preservation of information across the full polarized
boundary.

We define a unit-free coherent quantum core generated by multiplicative
wiring, unitary gates on positive qudit registers, and contextual
coherent control, and prove that every one of its morphisms is
essentially unitary. The framework realizes an applied coherent quantum switch
and the unitary stages of equal-ratio one-slot supermap dilations with
explicit memory. Complete proofs appear in the full version of this
paper.
\end{abstract}

\section{Introduction}
\label{sec:intro}

Since the foundational work of Abramsky and
Coecke~\cite{abramskyCoecke2004}, Categorical Quantum Mechanics has
shown that tensorial composition, duality, and entanglement fit within
dagger compact closed categories~\cite{Selinger2007CPM}. Adding a
symmetric monoidal sum accommodates classical interfaces and
measurement~\cite{CoeckeKissinger2017}. String diagrams give a
representation-free account of process composition, independent of
any Hilbert-space model.

In finite-dimensional closed-system quantum mechanics, reversible
dynamics is represented by unitary operators: an evolution $f:A\to A$
satisfies $f^\dagger f=ff^\dagger=\id_A$. Higher-order processes,
however, take processes rather than states as inputs. The quantum
switch~\cite{chiribella2013quantum}, for instance, coherently
superposes the two causal orders $f\circ g$ and $g\circ f$ of
evolutions $f,g:A\to A$ under quantum control; it is physically
realizable and reversible, yet is not a first-order unitary
endomorphism because its inputs are processes. This raises the
question: \emph{What boundary invariant extends first-order unitarity
to higher-order typed interfaces?}

We work from below. The base is the free compact closed category of
Kelly--Laplaza (KL)~\cite{coherence}, in the concrete combinatorial
presentation of Abramsky~\cite{AbramskyAS}, which separates
permutations, loops, and polarities. Its finite-coproduct completion
$\mathrm{Cop}$ records distinguishable layouts. Rather than mediate
distributivity by coherence isomorphisms~\cite{Laplaza72}, we work in
distributed normal form: objects are finite disjoint unions of
interfaces, and tensor distributes over sum at the object level. The
$\CC$-linear matrix/biproduct completion is the carrier category $\Perm(\CC)$,
a dagger compact closed category with finite biproducts, supplying
superposition and quantum amplitudes
(Section~\ref{sec:structural-core}).

A KL morphism is read on the two polarities of its compact-name
boundary. A basis linking contributes its permutation matrix; linear
combinations contribute the corresponding complex matrix, with closed
loop factors already in the coefficients. Essential unitarity (EU)
requires the resulting boundary matrix to be two-sided unitary
(Definition~\ref{def:EU}). On positive qudit registers this is ordinary
matrix unitarity. For qudits carrying multiplicative payloads, the full
matrix retains both the qudit-family address and the KL port
(Section~\ref{sec:eu-predicate}).

The source category $\QC$ (Definition~\ref{def:basic-QC}) has raw
types generated from positive qudit literals and their duals by
$\otimes$ and $\parr$. Its morphisms are generated by the unit-free no-Mix fragment of
multiplicative linear logic (MLL), unitary gates on positive qudit registers,
and contextual qudit control. A controlled family may share an
arbitrary higher-order input, but has a common positive output and
retains one selector port throughout. The two carrier counterexamples
of Section~\ref{subsec:QC-why-coherence} explain these restrictions.

For a source type \(X\), let \(\mathsf A_X^+\) and
\(\mathsf A_X^-\) collect its positive and negative boundary
coordinates, respectively. For \(h:A\to B\), compact naming exposes
the boundary \((A^*,B)\). Its fixed polarization determines the
incoming and outgoing axes, and typed un-naming gives
\[
 M(h):
 \mathsf A_A^+\otimes\mathsf A_B^-
 \longrightarrow
 \mathsf A_A^-\otimes\mathsf A_B^+.
\]
Our main theorem proves both Gram identities for \(M(h)\). Its
fixed-width representation is therefore a unitary numerical
qudit--MLL matrix, hence \(h\) is essentially unitary
(Theorem~\ref{thm:calculus-EU}).

The calculus also contains the applied contextual quantum switch of
Section~\ref{subsec:contextual-switch-v2}. Function types support
resource-linear one-slot transformations by pre- and postcomposition,
tensoring with constants, and synchronized qudit control. Finally, the
calculus realizes the unitary stages of deterministic one-slot
supermap dilations whenever the input/output dimension ratios agree
and the memory remains an explicit boundary wire
(Section~\ref{sec:expressiveness}). The recovery identity is stated in
external Hilbert-space/channel semantics; measurement, partial trace,
and mixed-state semantics are left to future work.

\subsection{Related Work}\label{sec:related-work}

\noindent\emph{Categorical quantum mechanics.}
Diagrammatic calculi such as the ZX-calculus~\cite{ZX} provide rewrite
systems for linear maps on qubits; dagger Frobenius algebras capture
observables and classical interfaces~\cite{Coecke2013}; and the CPM
construction~\cite{Selinger2007CPM} shows how mixed-state quantum
mechanics and completely positive maps arise canonically from dagger
compact structure.

\noindent\emph{Higher-order quantum processes.}
The study of \emph{supermaps}---transformations on quantum
operations---originates
with~\cite{Chiribella2008Supermaps,Chiribella2008,Chiribella2009},
which characterize deterministic one-slot supermaps and give
sequential realizations by pre- and post-isometries connected by a
memory system, followed by partial trace over the final memory.
Subsequent work studies causal and indefinite-order
processes~\cite{Oreshkov2012,A2014,KissingerUijlen2019,WilsonChiribellaKissingerLocality}
and higher-order-process frameworks based on BV logic,
closed symmetric monoidal categories, enriched monoidal categories, and
strong profunctors~\cite{simmons2022higherorder,Wilson2021HOPT,Wilson2022Framework,HeffordWilson2024}.
These approaches characterize higher-order admissibility relative to
an underlying first-order process theory through operational, causal,
or structural conditions. Two concurrent language-level accounts place
higher-order quantum control in causal process semantics. Hirata and
Tsukada~\cite{HirataTsukada2026} develop a pure, measurement-free
higher-order calculus with quantum conditionals and a causality type
system, interpreted in $\mathbf{CausHilb}$.
Barsse, P\'echoux, and Perdrix~\cite{BarssePechouxPerdrix2026}
extend coherent control to measurement and arbitrary quantum channels,
with denotational semantics in $\mathbf{Caus}[\mathbf{CPM}]$. At the
level of physical operations, $\QC$ isolates the qudit-controlled,
essentially unitary boundary fragment of this shared design space, leaving
measurement and discarding external.

\noindent\emph{The present approach.}
The construction begins with
Kelly--Laplaza compact-closed syntax and its complex-linear
finite-biproduct carrier. The source category $\QC$ adds positive
qudit-register gates and contextual qudit control to the unit-free
multiplicative base. Correctness is expressed by essential unitarity of
the full qudit--MLL boundary matrix of each $\QC$-morphism.

Our resource-linear contexts provide a unitary analogue of part of the
usual supermap
picture~\cite{WilsonChiribellaKissingerLocality,HeffordWilson2024}:
the distinguished hole is used once, substitution remains in $\QC$,
and its ambient carrier extension is $\CC$-linear.
The unit-free multiplicative structure of $\QC$ determines a symmetric
polycategory representable on nonempty
lists~\cite{Wilson2026HigherOrderCircuits,CockettSeely1997WDC}, and
resource-linear contexts give a source-level analogue of the
single-hole local-application fragment of Wilson and
Chiribella~\cite{WilsonChiribella2026MonoidalSU}. Equal input/output
dimension ratios permit padding the ancillary spaces and extending the
pre- and post-isometries to unitaries while memory wires remain
explicit.

\noindent\emph{Structure of the paper.}
Section~\ref{sec:orientation} introduces the boundary viewpoint and
essential unitarity.
Section~\ref{sec:structural-core} constructs the finite-coproduct
precursor $\mathrm{Cop}$ and the carrier $\Perm(\CC)$.
Sections~\ref{sec:boundary-transport} and~\ref{sec:eu-predicate}
develop KL boundary matrices, qudit--MLL matrices, and EU.
Section~\ref{sec:QC} defines $\QC$ and states the
essential-unitarity theorem.
Section~\ref{sec:expressiveness} develops resource-linear contexts
and equal-ratio one-slot dilation stages with explicit memory.
\extendedonly{The appendices provide the carrier, boundary-execution,
and motivating-witness calculations used in the main text.}

\section{Essential unitarity, informally}
\label{sec:orientation}

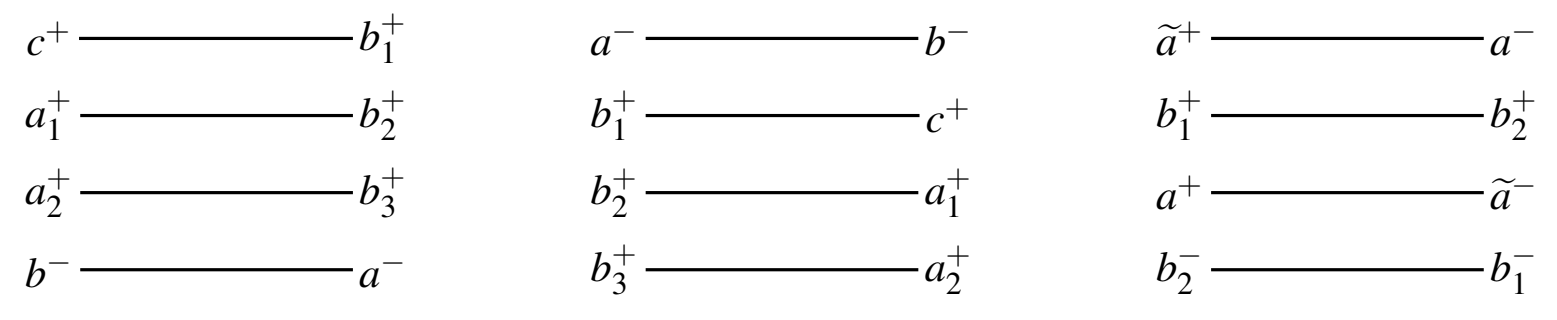
\begin{figure}[htbp]
\centering
\begin{tikzpicture}[scale=0.85,
  port/.style={font=\scriptsize, inner sep=1pt},
  wire/.style={semithick},
]
  \begin{scope}[local bounding box=klview]
    \node[port] (s1) at (0,1.8)   {$c^{+}$};
    \node[port] (s2) at (0,1.2)   {$a_1^{+}$};
    \node[port] (s3) at (0,0.6)   {$a_2^{+}$};
    \node[port] (s4) at (0,0)     {$b^{-}$};
    \node[port] (t1) at (2.6,1.8) {$b_1^{+}$};
    \node[port] (t2) at (2.6,1.2) {$b_2^{+}$};
    \node[port] (t3) at (2.6,0.6) {$b_3^{+}$};
    \node[port] (t4) at (2.6,0)   {$a^{-}$};
    \draw[wire] (s1) -- (t1);
    \draw[wire] (s2) -- (t2);
    \draw[wire] (s3) -- (t3);
    \draw[wire] (s4) -- (t4);
  \end{scope}
  \node[below=3pt of klview.south, font=\scriptsize]
    {(a)\ KL view};

  \begin{scope}[xshift=4.4cm, local bounding box=bview]
    \node[port] (p1) at (0,1.8)   {$a^{-}$};
    \node[port] (p2) at (0,1.2)   {$b_1^{+}$};
    \node[port] (p3) at (0,0.6)   {$b_2^{+}$};
    \node[port] (p4) at (0,0)     {$b_3^{+}$};
    \node[port] (n1) at (2.6,1.8) {$b^{-}$};
    \node[port] (n2) at (2.6,1.2) {$c^{+}$};
    \node[port] (n3) at (2.6,0.6) {$a_1^{+}$};
    \node[port] (n4) at (2.6,0)   {$a_2^{+}$};
    \draw[wire] (p1) -- (n1);
    \draw[wire] (p2) -- (n2);
    \draw[wire] (p3) -- (n3);
    \draw[wire] (p4) -- (n4);
  \end{scope}
  \node[below=3pt of bview.south, font=\scriptsize]
    {(b)\ boundary view};

  \begin{scope}[xshift=8.8cm, local bounding box=evview]
    \node[port] (e1) at (0,1.8)   {$\widetilde{a}^{+}$};
    \node[port] (e2) at (0,1.2)   {$b_1^{+}$};
    \node[port] (e3) at (0,0.6)   {$a^{+}$};
    \node[port] (e4) at (0,0)     {$b_2^{-}$};
    \node[port] (f1) at (2.6,1.8) {$a^{-}$};
    \node[port] (f2) at (2.6,1.2) {$b_2^{+}$};
    \node[port] (f3) at (2.6,0.6) {$\widetilde{a}^{-}$};
    \node[port] (f4) at (2.6,0)   {$b_1^{-}$};
    \draw[wire] (e1) -- (f1);
    \draw[wire] (e2) -- (f2);
    \draw[wire] (e3) -- (f3);
    \draw[wire] (e4) -- (f4);
  \end{scope}
  \node[below=3pt of evview.south, font=\scriptsize]
    {(c)\ $\ev_{A,B}$};
\end{tikzpicture}

\caption{A KL matching (a), its polarized boundary view (b), and
\(\ev_{A,B}\) for \(A=B=(1,1)\) (c).}
\label{fig:kl-vs-boundary}
\end{figure}
The starting point is the Kelly--Laplaza picture of a
morphism~\cite{coherence}, in the abstract-scalar presentation of
Abramsky~\cite{AbramskyAS}. A KL interface is a signed pair
$A=(A^+,A^-)$, with $\Pos(A):=A^+$ and $\Neg(A):=A^-$. Duality swaps
the two components, while tensor takes their disjoint union componentwise.
A KL morphism
$f:A\to B$ is a port matching
$A^+\sqcup B^-\cong B^+\sqcup A^-$, possibly with closed internal
loops. Its geometry is a wiring diagram.

Take $C=(1,0)$, $A=(2,1)$, and $B=(3,1)$. Then
$C\otimes A=(3,1)$, so a morphism $C\otimes A\to B$ matches four ports
on each side (Figure~\ref{fig:kl-vs-boundary}(a)). A KL basis morphism
is a port matching. The carrier $\Perm(\CC)$ adds complex amplitudes
and finite-family indices; within the single-interface block considered
here, its morphisms are finite $\CC$-linear combinations of such
matchings.

The boundary view regroups the same external ports as
$P_{A,B}:=\Neg(A)\sqcup\Pos(B)=\Pos(A^*\otimes B)$ and
$N_{A,B}:=\Pos(A)\sqcup\Neg(B)=\Neg(A^*\otimes B)$. The KL matching is
then read from $N_{A,B}$ (incoming) to $P_{A,B}$ (outgoing); in the running
example, $|P_{C\otimes A,B}|=|N_{C\otimes A,B}|=4$
(Figure~\ref{fig:kl-vs-boundary}(b)). The compact-name object
$A^*\otimes B$ changes the boundary partition, not the wiring. For KL
interfaces $A,B$, the carrier evaluation
$\ev_{A,B}:(A^*\otimes B)\otimes A\to B$ contracts the $A^*$- and
$A$-ports and carries the $B$-ports through.
Evaluation is syntactically higher-order but diagrammatically
ordinary. For $A=B=(1,1)$ it is the four-wire matching of
Figure~\ref{fig:kl-vs-boundary}(c); tildes mark ports from the
$A^*$-factor, and subscripts distinguish the two occurrences of $B$.
Its higher-order character lies in the typed boundary partition, not
the wiring.
Currying preserves each typed Kelly--Laplaza component and its
coefficient under the canonical identification of compact-name ports
(Section~\ref{subsec:k-linear}). It
changes the boundary partition, not the underlying matching.
Let $M,N$ be signed KL interfaces. For a loop-free matching
$g:M\to N$, write $[g]$ for its carrier basis vector,
$\tau_g:N_{M,N}\xrightarrow{\cong}P_{M,N}$ for its boundary bijection,
and $\underline{\tau_g}$ for the corresponding permutation matrix. For
a finite set $S$, let $\CC[S]$ be the Hilbert space with orthonormal
basis $S$. Thus
\[
 F=\sum_g c_g[g]:M\longrightarrow N
\]
has boundary matrix
\[
 \mathsf B_{M,N}(F)
 :=\sum_g c_g\,\underline{\tau_g}:
 \CC[N_{M,N}]\longrightarrow\CC[P_{M,N}],
\]
where the coefficients already contain the evaluated closed-loop
factors. Essential unitarity requires this operator to be two-sided
unitary. It is a boundary condition, not the assertion
that $F$ is a dagger-unitary morphism of $\Perm(\CC)$; evaluation is
the basic example separating the two notions.

Finite-family indices introduce one further coordinate. For a positive
qudit register carrying an MLL payload, the full matrix keeps the qudit
address separate from the KL-port matching. In a fixed-width sector,
every payload row has the same number \(r\) of KL ports; \(i,j\) are
qudit addresses and \(g\) ranges over \(r\)-port matchings. The matrix
has the form
\[
 \sum_{j,i,g}c_{ji,g}\,
 \lvert j\rangle\!\langle i\rvert
 \otimes\underline{\tau_g}.
\]
Section~\ref{sec:eu-predicate} develops this qudit--MLL matrix and its
sequent-level version. For a source morphism $h:A\to B$, compact naming
of its carrier image exposes the boundary $(A^*,B)$. The type fixes the
incoming factors $\mathsf A_A^+\otimes\mathsf A_B^-$ and the outgoing
factors $\mathsf A_A^-\otimes\mathsf A_B^+$; un-naming along this
partition produces $M(h)$. Its fixed-width representation turns the
carrier Gram identities into ordinary unitarity of the full numerical
matrix $\mathbf M(h)$.

The two counterexamples of Section~\ref{subsec:QC-why-coherence}
motivate the qudit-control discipline of Section~\ref{sec:QC}. Section~\ref{sec:calculus} states the essential-unitarity theorem;
the full proof reduces mixed-intermediate composition to
pure-intermediate composition.\par\smallskip
\emph{Throughout, the test is the same: retain both the qudit address
and the polarized KL ports, form the full boundary matrix, and ask
whether it is unitary.}

\section{The Polarized Carrier Category \texorpdfstring{\(\Perm(\CC)\)}{Perm(C)}}
\label{sec:structural-core}

Three names are reserved throughout: \emph{source} is $\QC$,
\emph{carrier} is $\Perm(\CC)$, \emph{precursor} is $\mathrm{Cop}$.
Put $\Bbbk:=\CC[\delta]$, with $\delta$ a formal indeterminate
recording closed execution cycles.  We construct the polynomial
matrix/biproduct envelope $\Perm(\Bbbk)$ in three steps --- the Kelly--Laplaza
base $\mathbf{KL}$~\cite{AbramskyAS,coherence} (ports), its free
finite-coproduct completion $\mathrm{Cop}$ (finite-family indexing,
with the groupoid $\mathbf{FinBij}$ of finite sets and bijections
governing structural reindexings), and the
full $\Bbbk$-linear matrix/biproduct completion (amplitudes) --- and
then specialize to
the carrier $\Perm(\CC)$, on which essential unitarity will live,
with loop-free KL matrix units as its basis morphisms.
\emph{Notation:} $\mathbf{KL}$ and $\Perm(\CC)$ carry a single
tensor; when a formula tracks source polarity we write
$X\parr Y:=X\otimes Y$ and $f\parr g:=f\otimes g$, the
identification fixed by the De Morgan coherences.

\subsection{Kelly--Laplaza base and polarized interfaces}
\label{subsec:kl-polarized}

The free dagger compact closed category on one generating object is
described concretely by Kelly and
Laplaza~\cite{AbramskyAS,coherence}.  The generator produces the
atomic positive interface $(1,0)$; duality gives $(0,1)$; tensor
builds all signed interfaces
$(m,n)=(1,0)^{\otimes m}\otimes(0,1)^{\otimes n}$.  Thus an object
of the Kelly--Laplaza base $\mathbf{KL}$ is a signed finite set
$M=(M^+,M^-)$, and a morphism $f:M\to N$ is a polarized bijection
$f^\sharp:M^+\sqcup N^-\cong N^+\sqcup M^-$ equipped with a loop
count $\kappa\in\NN$.

Composition is path execution with cycle formation: gluing along the
shared boundary decomposes each connected component into either a
path, contributing to the composite wiring, or a closed cycle; if
$\ell(g,f)$ denotes the number of new cycles, then
\[
\kappa(g\circ f) \;=\; \kappa(f)+\kappa(g)+\ell(g,f).
\]
Associativity and the correctness of the loop accounting follow from
Kelly--Laplaza coherence~\cite{AbramskyAS,coherence}.  The dagger is
inverse wiring; duality is polarity reversal, $M^*:=(M^-,M^+)$; the
tensor unit is $\Unit:=(\varnothing,\varnothing)$, the empty
interface, arising as the nullary tensor.

\subsection{The finite-coproduct precursor
\texorpdfstring{$\mathrm{Cop}$}{Cop}}
\label{subsec:ambient-coproducts}

We lift from individual interfaces to finite indexed families,
obtaining genuine finite coproducts.

\begin{definition}[Objects in $\oplus/\otimes$ normal form]
\label{def:distributed-objects}
An object in $\oplus/\otimes$ \emph{normal form} is a finite indexed
family
\[
A \;=\; \bigoplus_{i\in I} M_i,
\qquad M_i\in\mathbf{KL},
\]
with $I$ possibly empty (the empty family is the initial
object~$0$).  In family normal form the tensor unit is the singleton
family containing the empty KL interface~$\Unit$; it is distinct from
the empty-family zero object~$0$.  Tensor and objectwise polarity
reversal of normal-form objects are
\[
\Bigl(\bigoplus_{i\in I}M_i\Bigr)\otimes
\Bigl(\bigoplus_{j\in J}N_j\Bigr)
:=
\bigoplus_{(i,j)\in I\times J}(M_i\otimes N_j),
\qquad
\Bigl(\bigoplus_{i\in I}M_i\Bigr)^{*}
:=\bigoplus_{i\in I}M_i^{*},
\]
where at the port level
$(M^+,M^-)\otimes(N^+,N^-)=(M^+\sqcup N^+,\,M^-\sqcup N^-)$.
\end{definition}

At the index level, $\oplus$ is the $+$ of $\mathbf{FinBij}$
(concatenation of families) and $\otimes$ is its $\times$ (bilinear
distribution). For $A=\bigoplus_{i\in I}M_i$ and
$\epsilon\in\{+,-\}$, put
\[
 \operatorname{Ports}^{\epsilon}(A)
 :=\bigsqcup_{i\in I}M_i^{\epsilon},
 \qquad
 \Pos(A):=\operatorname{Ports}^{+}(A),
 \qquad
 \Neg(A):=\operatorname{Ports}^{-}(A).
\]
For a single interface,
$\operatorname{Ports}^{\epsilon}(M)=M^\epsilon$, extending the earlier
$\Pos/\Neg$ convention.

Every object of $\mathrm{Cop}$ and of its full $\Bbbk$-linear
matrix/biproduct completion is
in normal form, so $\otimes$ distributes over $\oplus$ by construction.
This indexing discipline prevents distributivity from being read as
port copying; see Section~\ref{sec:eu-predicate}.
\begin{definition}[The finite-coproduct precursor \(\mathrm{Cop}\)]
\label{def:cop-pm}
$\mathrm{Cop}:=\mathrm{Cop}(\mathbf{KL})$ has as objects the
normal-form objects of Definition~\ref{def:distributed-objects}, and
as morphisms $f:A\to B$, for $A=\bigoplus_{i\in I}M_i$ and
$B=\bigoplus_{j\in J}N_j$, pairs
$f=\bigl(\varphi,(f_i)_{i\in I}\bigr)$ with
\begin{enumerate}[label=\textup{(\roman*)},nosep]
\item $\varphi:I\to J$ a function;
\item $f_i:M_i\to N_{\varphi(i)}$ a Kelly--Laplaza morphism, for
each $i\in I$.
\end{enumerate}
Composition is componentwise:
$\bigl(\psi,(g_j)\bigr)\circ\bigl(\varphi,(f_i)\bigr)
 :=\bigl(\psi\varphi,\,(g_{\varphi(i)}\circ f_i)\bigr)$.
\end{definition}

The canonical injections and cotuplings exhibit $\mathrm{Cop}$ as
the free finite-coproduct completion of~$\mathbf{KL}$.

\subsection{The full \texorpdfstring{$\Bbbk$}{k}-linear matrix/biproduct completion \texorpdfstring{$\Perm(\Bbbk)$}{Perm(k)}}
\label{subsec:k-linear}

Equip $\Bbbk=\CC[\delta]$ with the involution that conjugates complex
coefficients and fixes $\delta$. Definition~\ref{def:perm-complex}
below specializes $\delta\mapsto d$ along $\CC[\delta]\to\CC$, for a
chosen $d\in\RR_{>0}\setminus\{1\}$.\footnote{The choice $d\neq1$ makes
closed trace loops detectable on a nonempty residual boundary.
The full version shows that every correct
unit-free no-Mix multiplicative component used by the source calculus
is a coefficient-one matching with no detached execution component,
and hence contributes no power of $d$.}

\begin{definition}[Full $\Bbbk$-linear matrix/biproduct completion]
\label{def:perm-C}
Write $\mathbf{KL}^{0}(M,N)$ for the loop-free KL morphisms and $[f]$
for the basis vector of $f\in\mathbf{KL}^{0}(M,N)$. For
$A=\bigoplus_{i\in I}M_i$ and $B=\bigoplus_{j\in J}N_j$, set
\[
 \Perm(\Bbbk)(A,B)
 :=\mathrm{Mat}_{J\times I}
 \bigl(\Bbbk[\mathbf{KL}^{0}(M_i,N_j)]\bigr).
\]
Each entry is therefore a finite $\Bbbk$-linear combination of
loop-free matchings. For $f\in\mathbf{KL}^{0}(M,N)$ and
$g\in\mathbf{KL}^{0}(N,P)$, let $g\star f$ be the external-path
matching of their glued graph and let $\ell(g,f)$ be its number of
closed cycles. Define
\begin{equation}
 \label{eq:delta-twisted-basis-composition}
 [g]\circ[f]:=\delta^{\ell(g,f)}[g\star f].
\end{equation}
Extend this law $\Bbbk$-bilinearly to
$\mathcal L:=\Bbbk[\mathbf{KL}^{0}]$. Composition in $\Perm(\Bbbk)$
is matrix multiplication over $\mathcal L$. Dagger is conjugate transpose with
$[f]^\dagger=[f^\dagger]$ and $\bar\delta=\delta$. The linearization
of a general KL morphism is
\[
 \mathbf{KL}\longrightarrow\mathcal L,
 \qquad
 (f^\sharp,\kappa)\longmapsto\delta^{\kappa}[f^\sharp].
\]
\end{definition}

An occupied Kelly--Laplaza component is \emph{coefficient-one} if it is
a single basis matching with scalar \(1\). On pure-axis summands this
matching is a Kelly--Laplaza basis isomorphism. A matrix is
coefficient-one if every occupied entry is coefficient-one.

Brackets are suppressed when no confusion can result.  The image of
$(\tau,\kappa)$ under the linearization is $\delta^{\kappa}[\tau]$:
the loop count contributes to the coefficient, not to the basis
label.

$\mathrm{Cop}\hookrightarrow\Perm(\Bbbk)$ on the same objects: its
image is the subcategory with exactly one nonzero entry per column,
that entry a single $\mathbf{KL}$ monomial $\delta^{\kappa}[\tau]$
supplied by the data of Definition~\ref{def:cop-pm}.  So
$\mathrm{Cop}$ pins the $\oplus$-indexing, and $\Perm(\Bbbk)$ adds
$\Bbbk$-linear combinations in each entry.

\extendedonly{%
For a pair of loop-free KL basis morphisms, the gluing can be computed
by a finite transient-path sum after its closed internal cycles have
been extracted.  No inverse over \(\Bbbk\) is used;
Appendix~\ref{app:structural-proofs} gives the basiswise path-execution
proof.%
}
\begin{proposition}[The loop-erased envelope]
\label{prop:well-defined-assoc}
The rule~\eqref{eq:delta-twisted-basis-composition} makes the
loop-erased $\Bbbk$-linear envelope
$\mathcal L=\Bbbk[\mathbf{KL}^{0}]$ a dagger compact closed category.
Its compact structure supplies the canonical Joyal--Street--Verity
trace, the linearization $\mathbf{KL}\to\mathcal L$ is functorial,
and matrix composition over $\mathcal L$ makes $\Perm(\Bbbk)$ a
dagger compact closed category with finite coproducts.
\end{proposition}

\begin{proof}
Appendix~\ref{app:structural-proofs} gives the direct componentwise
proof of the category laws, compact equations, functoriality of the
linearization, and the basis-level feedback calculation.
\end{proof}

Its tensor is inherited from $\mathbf{KL}$, while its
finite-coproduct indexing is governed by $\mathbf{FinBij}$; every
object remains in the $\oplus/\otimes$ normal form of
Definition~\ref{def:distributed-objects}.

\paragraph*{Biproducts in $\Perm(\Bbbk)$.}
Since $\Perm(\Bbbk)$ is $\Bbbk$-linear with a zero object (the
empty family), its finite coproducts are biproducts by standard
semiadditivity.  The source calculus introduced later does not expose
biproduct injections or projections as source constructors; such maps
remain available for carrier-level matrix arguments in $\Perm(\CC)$.

A morphism $f:A\otimes B\to C$ and its curry
$\tilde f:A\to B^*\otimes C$ share the same underlying wiring:
currying merely repartitions boundary ports.
On basis morphisms this is the identity on polarized bijections;
it extends $\Bbbk$-linearly to all of $\Perm(\Bbbk)$.

\begin{definition}[$\Perm(\CC)$]
\label{def:perm-complex}
Fix $d\in\RR_{>0}\setminus\{1\}$ and let
$\operatorname{ev}_d:\CC[\delta]\to\CC$ be evaluation at
$\delta=d$.  Then $\Perm(\CC)$ is the coefficientwise specialization
of $\Perm(\Bbbk)$ along $\operatorname{ev}_d$: it has the same
objects, and its hom-spaces are obtained by applying
$\operatorname{ev}_d$ to every coefficient.
\end{definition}

Because $d$ is real, the evaluation preserves the involution;
consequently $\Perm(\CC)$ is a $\CC$-linear dagger compact closed
category with tensor unit $\Unit$ and finite biproducts $\oplus$.
By abuse of notation we continue to write $\delta$ for the chosen
scalar~$d$.  For a finite set $S$, write $\CC[S]$ for the
finite-dimensional Hilbert space with orthonormal basis $S$; this
set-linearization notation is distinct from the polynomial ring
$\CC[\delta]$.

Every morphism $F:A\to B$ of $\Perm(\CC)$, with
$A=\bigoplus_{i\in I}M_i$ and $B=\bigoplus_{j\in J}N_j$, is by
definition a $J\times I$ matrix
\[
  F=(F_{ji}),\qquad F_{ji}\in\CC[\mathbf{KL}^{0}(M_i,N_j)].
\]
Writing each entry uniquely as a finite collected sum of distinct
loop-free KL matchings,
\[
  F_{ji}=\sum_r \alpha_{ji,r}\,g_{ji,r},
  \qquad g_{ji,r}\in\mathbf{KL}^{0}(M_i,N_j),
\]
gives the \emph{collected KL matrix form} of $F$; each coefficient
$\alpha_{ji,r}$ already contains the evaluated powers of $d$
produced by closed cycles, so loop counts are not additional basis
data.

\begin{definition}[First-order objects]
\label{def:first-order}
An object $A=\bigoplus_{i\in I}M_i$ is \emph{first-order} if all its
summands are pure-sign of the same polarity: $M_i=(m_i,0)$ for all
$i$, or $M_i=(0,m_i)$ for all $i$.  A morphism is
\emph{first-order} if its domain and codomain are first-order of the
same polarity.  Write $\mathbf{H}_n := (1,0)^{\oplus n}$ for the
first-order object of \emph{size}~$n$.
\end{definition}

Fix either polarity.  In the positive sector there are no negative
ports (the negative sector is its dual/opposite), so no new closed cycles
form under composition: the $\delta$-twist disappears, and
composition is the usual matrix-sum formula with entry
multiplication given by composition of loop-free KL matchings.  For
the one-port objects $\mathbf H_n$ the matching spaces are
one-dimensional, and this reduces to ordinary complex matrix
multiplication.

Section~\ref{sec:boundary-transport} first assigns boundary matrices
to morphisms between individual KL interfaces, then incorporates
finite-family qudit addresses in fixed-width qudit--MLL matrices and
uniform sequent matrices. The $\oplus/\otimes$ normal form and KL
matrix-unit basis make each stage unambiguous.

\section{Kelly--Laplaza Boundaries and Essential Unitarity}
\label{sec:boundary-transport}

Boundary transport reads a Kelly--Laplaza morphism \(M\to N\) on the
two polarities of its compact-name interface \(M^*\otimes N\). A basis
morphism contributes the permutation matrix of its axiom-link matching;
linear combinations contribute the corresponding \(\CC\)-linear
combination, with evaluated loop factors already present in the
coefficients. The type repartitions the boundary; the wiring is
unchanged. On the one-port registers \(\mathbf H_n\), this
recovers ordinary complex matrices. Qudits tensored with multiplicative
payloads will give the full matrices used later.

\subsection{KL boundary matrices}
\label{subsec:boundary-ports}

\begin{definition}[KL boundary matching]
\label{def:boundary-objects}
\label{def:boundary-bijection}
For signed KL interfaces \(M,N\), put
\(P_{M,N}:=\Neg(M)\sqcup\Pos(N)\) and
\(N_{M,N}:=\Pos(M)\sqcup\Neg(N)\).
If \(g:M\to N\) has polarized wiring
\(g^\sharp:N_{M,N}\xrightarrow{\cong}\Pos(N)\sqcup\Neg(M)\),
let \(\chi_{M,N}\) be the canonical swap onto \(P_{M,N}\), and set
\(\tau_g:=\chi_{M,N}g^\sharp:N_{M,N}\xrightarrow{\cong}P_{M,N}\).
Write
\(\underline{\tau_g}:\CC[N_{M,N}]\to\CC[P_{M,N}]\)
for its permutation matrix.
\end{definition}

We read the boundary covariantly, from \(\CC[N_{M,N}]\) to
\(\CC[P_{M,N}]\): \(g^\sharp\) is used without inversion, and its
codomain is placed in the stated \(P\)-order.

The map \(\tau_g\) is the same axiom-link wiring reorganized around
\(M^*\otimes N\). It is therefore square even when \(M\) and \(N\)
have different shapes. In particular, evaluation has differently
shaped source and target but a boundary consisting of straight wires.
Currying likewise preserves each component matching under the canonical
identification of compact-name ports
(Section~\ref{subsec:k-linear}).

Every \(F:M\to N\) in \(\Perm(\CC)\) has a unique KL-basis
expansion \(F=\sum_g c_g[g]\). Its coefficients already include all
evaluated loop factors.

\begin{definition}[KL boundary matrix]
\label{def:KL-boundary-matrix}
Define
\[
 \mathsf B_{M,N}(F)
 :=\sum_g c_g\,\underline{\tau_g}:
 \CC[N_{M,N}]\longrightarrow\CC[P_{M,N}].
\]
\end{definition}

\begin{definition}[Essential unitarity]
\label{def:EU}
A morphism \(F:M\to N\) is \emph{essentially unitary} if
\(\mathsf B_{M,N}(F)^\dagger\mathsf B_{M,N}(F)=I_{\CC[N_{M,N}]}\)
and
\(\mathsf B_{M,N}(F)\mathsf B_{M,N}(F)^\dagger=I_{\CC[P_{M,N}]}\).
Write \(\operatorname{EU}(F)\) for this condition.
\end{definition}

Every loop-free KL basis morphism is EU. A component with loop-free
part \(g\) and loop count \(\kappa\) has boundary matrix
\(d^\kappa\underline{\tau_g}\); on a nonempty boundary it is EU exactly
when \(\kappa=0\). The empty boundary cannot detect scalars.

This boundary condition is distinct from carrier dagger-unitarity. For
nonempty \(M\), the evaluation component
\[
 \ev_{M,N}:(M^*\otimes N)\otimes M\longrightarrow N
\]
is EU because its boundary matrix is a permutation, but
\[
 \ev_{M,N}\ev_{M,N}^\dagger
 =d^{|M^+|+|M^-|}\id_N\ne\id_N
\]
in \(\Perm(\CC)\).

The boundary representation also need not be injective on linear
combinations. For \(R_3:=(3,0)\), put
\[
 K:=\sum_{\sigma\in S_3}\operatorname{sgn}(\sigma)[\sigma].
\]
Then \(K\ne0\), whereas
\(\mathsf B_{R_3,R_3}(K)=0\): at each matrix entry, one even and one
odd permutation cancel. Hence \(\id_{R_3}+\lambda K\) is EU for every
\(\lambda\in\CC\). Since \(K^\dagger=K\) and \(K^2=6K\),
\[
 (\id_{R_3}+\lambda K)^\dagger(\id_{R_3}+\lambda K)
 =\id_{R_3}
  +(\lambda+\bar\lambda+6|\lambda|^2)K,
\]
so it is generally not carrier dagger-unitary. Thus EU records the
induced action on the polarized port space, not the entire carrier
element.

\subsection{Execution and loop coefficients}
\label{subsec:composition-law}

Because the boundary retains external paths and absorbs cycles into
coefficients, composition is governed by KL execution.

\begin{proposition}[KL execution]
\label{prop:boundary-execution-gluing}
For morphisms \(f:M\to N\) and \(g:N\to P\) in \(\mathbf{KL}\),
the external paths
determine \(\tau_{g\circ f}\), while
\(\kappa(g\circ f)=\kappa(f)+\kappa(g)+\ell(g,f)\),
where \(\ell(g,f)\) is the number of new internal cycles.
\end{proposition}

This is the paths-and-cycles decomposition of
\cite[Proposition~1]{AbramskyAS}; the port calculation appears in
the full version of this paper. For mixed interfaces it is a
geometric execution law, not multiplication of the square boundary
matrices: the output partition of \(f\) need not be the input partition
of \(g\).

\begin{lemma}[Dagger]
\label{lem:KL-boundary-dagger}
For every \(F:M\to N\),
\(\operatorname{EU}(F^\dagger)\iff\operatorname{EU}(F)\).
\end{lemma}

\begin{proof}
On basis morphisms, dagger inverts the matching and exchanges the two
boundary orders. These exchanges are unitary permutation matrices;
conjugating coefficients gives the two Gram identities.
\end{proof}

Cup--cap loops break closure: for nonempty \(M\), the unit
\(\eta_M\) and the composable counit
\(\epsilon_{M^*}\) are loop-free and hence EU, whereas
\[
 s_M:=\epsilon_{M^*}\eta_M
 =d^{|M^+|+|M^-|}\id_\Unit.
\]
For every nonempty \(X\),
\[
 (\epsilon_{M^*}\otimes\id_X)
 (\eta_M\otimes\id_X)
 =s_M\otimes\id_X
\]
multiplies \(\mathsf B_{X,X}(\id_X)\) by
\(d^{|M^+|+|M^-|}\).
Thus EU is not closed under composition or tensor in \(\Perm(\CC)\).

\section{Qudits, MLL Payloads, and Full Matrices}
\label{sec:eu-predicate}

On a pure-sign intermediate, KL execution becomes ordinary matrix
composition. Naming and polarized un-naming place a qudit with an
arbitrary multiplicative payload on such pure-positive axes; this
assignment is not functorial for arbitrary composition through a mixed
payload. These two steps give the full matrix used by the source
calculus.

\subsection{Positive qudit matrices}

For \(m,n\ge1\), write \([n]=\{1,\ldots,n\}\), and similarly
for \([m]\). Recall that
\(\mathbf H_n=(1,0)^{\oplus n}\) and that \(\CC[S]\) has orthonormal
basis \(S\). A morphism
\(U:\mathbf H_n\to\mathbf H_m\) is therefore its ordinary coefficient
matrix
\[
 U=\sum_{j,i}U_{ji}e_{ji}:\CC[n]\longrightarrow\CC[m],
 \qquad
 e_{ji}:=\lvert j\rangle\!\langle i\rvert.
\]
Carrier dagger-unitarity and ordinary matrix unitarity coincide here.

The same formula retains a nontrivial KL matching at each matrix entry.
Let \(R_r:=(r,0)\) for \(r\ge0\). Fix \(r\ge1\), and write
\[
 G_{ji}=\sum_{g\in\mathbf{KL}^{0}(R_r,R_r)}c_{ji,g}[g]
\]
for \(G:R_r^{\oplus n}\to R_r^{\oplus m}\).

\begin{definition}[Fixed-width qudit--KL matrix]
\label{def:first-order-boundary-matrix}
Define
\[
 \mathsf B_r(G)
 :=\sum_{j,i,g}c_{ji,g}\,
 e_{ji}\otimes\underline{\tau_g}:
 \CC[n]\otimes\CC[r]\longrightarrow
 \CC[m]\otimes\CC[r].
\]
\end{definition}

The two coordinates have different roles: \(i,j\) are qudit indices,
while \(\underline{\tau_g}\) acts on the \(r\) KL ports.

In one pure-sign sector, normal form keeps biproduct-family indices
separate from KL ports. Carrier distributivity only reindexes the
families; its KL components are identity matchings, so its matrix is a
unitary permutation. No port is copied.

\begin{proposition}[Fixed-width representation]
\label{prop:first-order-representation}
For composable \(G,H\) in the fixed \(r\)-sector,
\[
 \mathsf B_r(\id)=I,\qquad
 \mathsf B_r(HG)=\mathsf B_r(H)\mathsf B_r(G),\qquad
 \mathsf B_r(G^\dagger)=\mathsf B_r(G)^\dagger.
\]
\end{proposition}

\begin{proof}
A pure-positive intermediate has no opposite-polarity ports, so
execution creates no cycle. KL composition is ordinary permutation
composition, and the result is block-matrix multiplication.
\end{proof}

For \(r=1\), the matching space is one-dimensional and
\(\mathsf B_1(U)=U\).

\subsection{Qudits with MLL payloads}

For \(f:U\to V\) and \(a:\Unit\to U^*\otimes V\), use
\[
 \operatorname{name}(f):=(\id_{U^*}\otimes f)\eta_U,
 \qquad
 \operatorname{unname}(a):=
 (\epsilon_U\otimes\id_V)(\id_U\otimes a),
\]
where \(\eta_U:\Unit\to U^*\otimes U\) and
\(\epsilon_U:U\otimes U^*\to\Unit\). Fixed associativity and unit
coherence are suppressed; name and unname are inverse.

Let \(M=(M^+,M^-)\) and \(N=(N^+,N^-)\) be carrier KL
interfaces. For \(Y\in\{M,N\}\), put
\[
 Y_+:=(Y^+,\varnothing),
 \qquad
 Y_-:=(Y^-,\varnothing).
\]
These pure-positive interfaces give the canonical coefficient-one
factorization
\[
 Y\cong Y_-^*\otimes Y_+.
\]
Consequently, for
\[
 A:=\mathbf H_n\otimes M,
 \qquad
 B:=\mathbf H_m\otimes N,
\]
there are canonical displays
\[
 A\cong M_-^*\otimes(\mathbf H_n\otimes M_+),
 \qquad
 B\cong N_-^*\otimes(\mathbf H_m\otimes N_+).
\]

These displays explain the full qudit--MLL matrix. They move the qudit
index to the positive axis before the compact name is uncurried.

Let
\[
 \zeta_{M,N}^{n,m}:
 A^*\otimes B\xrightarrow{\cong}
 \bigl(\mathbf H_n\otimes(M_+\otimes N_-)\bigr)^*
 \otimes
 \bigl(\mathbf H_m\otimes(M_-\otimes N_+)\bigr)
\]
be the canonical coefficient-one compact-structural isomorphism
induced by duality, symmetry, and associativity.

\begin{definition}[Qudit--MLL carrier matrix]
\label{def:qudit-mll-carrier-matrix}
For \(F:A\to B\), define
\[
 \mathcal Q_{M,N}^{n,m}(F)
 :=
 \operatorname{unname}\!\left(
 \zeta_{M,N}^{n,m}
 \operatorname{name}_{\Perm(\CC)}(F)\right).
\]
This is a positive first-order carrier morphism from
\(\mathbf H_n\otimes(M_+\otimes N_-)\) to
\(\mathbf H_m\otimes(M_-\otimes N_+)\).
\end{definition}

This type supports a KL basis component precisely when
\[
 1+|M^+|+|N^-|=1+|M^-|+|N^+|.
\]
Assume this balance and write the common value as \(r\).
The extra \(1\) is the qudit's own KL port. After distribution,
\(\mathcal Q_{M,N}^{n,m}(F)\) lies in the fixed-width sector
\(R_r^{\oplus n}\to R_r^{\oplus m}\). Its full numerical matrix is
\[
 \mathbf Q_{M,N}^{n,m}(F)
 :=\mathsf B_r\!\left(\mathcal Q_{M,N}^{n,m}(F)\right):
 \CC[n]\otimes\CC[r]\longrightarrow\CC[m]\otimes\CC[r].
\]
For general linear combinations in this fixed block format,
unitarity is a condition on the full block matrix; unitary
constituent matchings alone do not suffice.

\subsection{Uniform qudit--MLL boundaries}

The preceding construction iterates over any nonempty one-sided
qudit--MLL boundary.  For a carrier formula \(X\) generated from
\(\mathbf H_n\), \(\mathbf H_n^*\), \(\otimes\), and \(\parr\), collect
its qudit rows by polarity:
\[
 \begin{aligned}
 \mathsf A_{\mathbf H_n}^-&:=\Unit,
 &\mathsf A_{\mathbf H_n}^+&:=\mathbf H_n,\\
 \mathsf A_{\mathbf H_n^*}^-&:=\mathbf H_n,
 &\mathsf A_{\mathbf H_n^*}^+&:=\Unit,\\
 \mathsf A_{X\diamond Y}^{\epsilon}
 &:=\mathsf A_X^{\epsilon}\otimes\mathsf A_Y^{\epsilon}
 &&(\diamond\in\{\otimes,\parr\},\ \epsilon\in\{-,+\}).
 \end{aligned}
\]
For a sequent \(\Sigma=(X_1,\ldots,X_s)\), tensor the corresponding
factors and put \(\mathsf A_\varnothing^\pm:=\Unit\).  Write
\(|\Sigma|:=X_1\otimes\cdots\otimes X_s\) for its carrier object
and \(|\varnothing|:=\Unit\).  Each polarized factor is pure positive.
Its uniform-row display is a fixed coefficient-one dagger
isomorphism:
\[
 d_\Sigma^\epsilon:
 \mathsf A_\Sigma^\epsilon
 \xrightarrow{\cong}
 R_{k_\Sigma^\epsilon}^{\oplus N_\Sigma^\epsilon}.
\]
Here \(k_\Sigma^\epsilon\) is its port width and
\(N_\Sigma^\epsilon\) its number of qudit-address rows; for the empty
factor these are \(0\) and \(1\).

The fixed compact coherence gives a coefficient-one dagger isomorphism
\[
 \zeta_\Sigma:
 |\Sigma|\xrightarrow{\cong}
 (\mathsf A_\Sigma^-)^*\otimes\mathsf A_\Sigma^+.
\]
This is the preceding qudit--MLL polarization applied simultaneously
to all literals of the boundary.

\begin{definition}[Qudit--MLL sequent matrix]
\label{def:qudit-mll-sequent-matrix}
For \(a:\Unit\to|\Sigma|\), define
\[
 \mathcal Q_\Sigma(a)
 :=\operatorname{unname}(\zeta_\Sigma a):
 \mathsf A_\Sigma^-\longrightarrow\mathsf A_\Sigma^+.
\]
If \(k_\Sigma^-=k_\Sigma^+=r\geq1\), its full numerical realization is
\[
 \mathbf Q_\Sigma(a)
 :=\mathsf B_r\!\left(
 d_\Sigma^+\mathcal Q_\Sigma(a)(d_\Sigma^-)^{-1}
 \right).
\]
\end{definition}

\begin{definition}[Qudit--MLL essential unitarity]
\label{def:qudit-mll-EU}
For a morphism
\(F:\mathbf H_n\otimes M\to\mathbf H_m\otimes N\) satisfying
\[
 1+|M^+|+|N^-|=1+|M^-|+|N^+|,
\]
qudit--MLL essential unitarity means that
\(\mathbf Q_{M,N}^{n,m}(F)\) is a two-sided unitary. For a sequent
point \(a:\Unit\to|\Sigma|\) with
\(k_\Sigma^-=k_\Sigma^+\geq1\), it means that
\(\mathbf Q_\Sigma(a)\) is a two-sided unitary.
\end{definition}

Carrier dagger-unitarity of \(\mathcal Q_\Sigma(a)\) implies the
sequent condition by Proposition~\ref{prop:first-order-representation}.

\section{The Coherent Syntax}
\label{sec:QC}

This section defines the source category \(\QC\), its carrier
interpretation, its unitary gates on positive qudit registers, and its
contextual control constructor. The qudit--MLL matrix interface is
developed in Section~\ref{sec:eu-predicate}; essential unitarity is
stated in Section~\ref{sec:calculus}.

\subsection{Why Qudit Control}
\label{subsec:QC-why-coherence}

Two carrier counterexamples in \(\Perm(\CC)\) motivate the source
restrictions below; the calculations appear in the full version.

For the first, put \(Q:=(1,0)=\mathbf H_1\), \(F:=Q^*\otimes Q\),
and \(M:=F\otimes Q\), and let \(\bar w:M\to M\) be the
coefficient-one multiplicative basis morphism with links
\(a\mapsto b'\), \(b\mapsto c\), \(c'\mapsto a'\), where
\(a,b\) are the positive function-output and payload ports, \(c\)
is the negative function-input port, and primes mark the target copy.
Let \(X:M\oplus M\to M\oplus M\) exchange the two summands, put
\(L:=\id_M\oplus\bar w\), and write \(\mathsf C(K)\) for the
full qudit--MLL matrix of \(K:M\oplus M\to M\oplus M\) under the
canonical identification \(\mathbf H_2\otimes F\cong M\oplus M\).
For \(\theta\notin\tfrac{\pi}{2}\mathbb Z\), the three factors of
\[
 h:=L\circ
 \bigl(\cos\theta\,\id_{M\oplus M}+\ii\sin\theta\,X\bigr)
 \circ L
\]
all have unitary full matrices, yet
\(\mathsf C(h)\notin\mathsf U(6)\): the composite is not
essentially unitary. Arbitrary carrier-\(\oplus\) control is
therefore not stable over a mixed payload, motivating explicit
selector retention; the contextual constructor below also requires a
positive common output, ensuring the pure-axis block form used by the
compositionality and normalization arguments.

For the second, put \(Y:=Q\otimes Q^*\) and \(Z:=Y\otimes Y\),
let \(\sigma:Z\to Z\) exchange the two \(Y\)-factors, and let
\(\bar q:Z\to Y\) be a suitable coefficient-one multiplicative
basis morphism. Although
\(\cos\theta\,\id_Z+\ii\sin\theta\,\sigma\) is carrier
dagger-unitary, for \(\theta\notin\tfrac{\pi}{2}\mathbb Z\) the
composite
\(\bar q\circ(\cos\theta\,\id_Z+\ii\sin\theta\,\sigma)\)
is not essentially unitary: composition through the mixed interface
\(Z\) destroys boundary unitarity. Primitive unitaries are therefore
admitted on positive qudit registers, not on arbitrary multiplicative
interfaces.

\subsection{Raw Types and Qudit Registers}
\label{subsec:QC-objects}

\begin{definition}[Raw types]
\label{def:QC-basic-types}
For \(n\geq1\), let \(\mathbf R_n\) be the positive \(n\)-level qudit
literal and put \(q:=\mathbf R_1\). Raw types are generated by
\(A::=\mathbf R_n\mid\mathbf R_n^*\mid A\otimes B\mid A\parr B\).
Their strict De Morgan duality is
\((\mathbf R_n^*)^*=\mathbf R_n\),
\((A\otimes B)^*=A^*\parr B^*\), and
\((A\parr B)^*=A^*\otimes B^*\).
The polarity set is determined by
\(\operatorname{pol}(\mathbf R_n)=\{+\}\),
\(\operatorname{pol}(\mathbf R_n^*)=\{-\}\), and
\(\operatorname{pol}(A\otimes B)=\operatorname{pol}(A\parr B)=\operatorname{pol}(A)\cup\operatorname{pol}(B)\).
A type is \emph{positive}, \emph{negative}, or \emph{mixed} according
as its polarity set is \(\{+\}\), \(\{-\}\), or \(\{+,-\}\).
A positive or negative type is \emph{pure}.
\end{definition}

The source grammar is unit-free: it has no multiplicative-unit type or
nullary compact constructor. The carrier category \(\Perm(\CC)\)
remains compact and unital. With its compact \(*\)-autonomous
structure, both source multiplicatives are interpreted by tensor.

\begin{definition}[Object assignment]
\label{def:QC-object-interpretation}
The raw-object assignment \(O\) is determined by
\[
 \begin{aligned}
 O(\mathbf R_n)&=\mathbf H_n=(1,0)^{\oplus n},
 & O(\mathbf R_n^*)&=\mathbf H_n^*,\\
 O(A\otimes B)&=O(A)\otimes O(B),
 & O(A\parr B)&=O(A)\otimes O(B).
 \end{aligned}
\]
\end{definition}

The source qudit is not a multiplicative unit: every summand of
\(\mathbf H_n\) carries one positive KL port. In particular,
\(\mathbf R_m\otimes\mathbf R_n\) and \(\mathbf R_{mn}\) have equally
many carrier rows but different multiplicative interfaces.

Write \([n]:=\{1,\ldots,n\}\). For positive types, define the address
set \(I_P\) and port width \(k_P\) recursively by
\[
 \begin{aligned}
 I_{\mathbf R_n}&:=[n],
 &
 k_{\mathbf R_n}&:=1,\\
 I_{P\diamond S}&:=I_P\times I_S,
 &
 k_{P\diamond S}&:=k_P+k_S
 \qquad(\diamond\in\{\otimes,\parr\}),
 \end{aligned}
\]
and put \(N_P:=|I_P|\). The carrier \(\oplus/\otimes\) normal form
supplies the fixed lexicographic qudit-row display
\begin{equation}
 d_P:O(P)\xrightarrow{\cong}
 \bigoplus_{\mathbf i\in I_P}(k_P,0).
 \label{eq:QC-positive-display}
\end{equation}
Thus \(I_P\) records qudit addresses, whereas each row retains all
\(k_P\) multiplicative ports.

\begin{definition}[Positive-register gates]
\label{def:QC-register-unitaries}
Positive types \(P,Q\) are \emph{shape-compatible} when
\(N_P=N_Q\) and \(k_P=k_Q\). For every shape-compatible pair and every unitary
\(U:\CC[I_P]\to\CC[I_Q]\), introduce
\(u_U^{P,Q}:P\to Q\), subject to
\(u_I^{P,P}=\id_P\) and
\(u_V^{Q,R}\circ u_U^{P,Q}=u_{VU}^{P,R}\).
Negative register gates are their duals.
\end{definition}

Write \(u_U^P:=u_U^{P,P}\). For \(P=\mathbf R_n\), write
\(u_U:=u_U^{\mathbf R_n}\). Under the carrier assignment below,
\(u_U\) has ordinary coefficient matrix \(U\). Allowing \(P\) to be
compound supplies joint,
including entangling, unitaries on
\(\mathbf R_m\otimes\mathbf R_n\) without identifying that two-port
type with the one-port literal \(\mathbf R_{mn}\).

\subsection{The Category \texorpdfstring{\(\QC\)}{QC}}
\label{subsec:QC-category}

The remaining constructor realizes a synchronized family with one
qudit selector.  For a finite raw context
\(\Gamma=(A_1,\ldots,A_k)\), write
\[
 \Gamma\mathbin{\cdot}D
 :=A_1\otimes\cdots\otimes A_k\otimes D,
 \qquad
 \varnothing\mathbin{\cdot}D:=D,
\]
using the fixed association convention.

Let \(\mathsf{Obj}_{\mathrm u}\) be the formulas generated from the
qudit literals by \(\Unit,\bot,\otimes,\parr\), and strict De Morgan
duality, with \(\Unit^*=\bot\) and \(\bot^*=\Unit\).  Unit
occurrences remain visible in these formulas; the raw types are exactly
the unit-free ones.  We use the typed-component presentation
of~\cite{BCST1996}; the category-labelled construction
of~\cite{Hughes2012} is its ordinary-category special case.

\begin{definition}[Contextual qudit control]
\label{def:basic-admissible-coherent-tuple}
For \(X,Y\in\mathsf{Obj}_{\mathrm u}\), let
\(\mathsf{Term}(X,Y)\) be the least family of finite typed terms
generated by the unital symmetric \(*\)-autonomous term formers,
the positive-register gates, and
\begin{equation}
 \frac{f_i:\Gamma\mathbin{\cdot}D\to P\quad(i\in[n])}
      {\operatorname{ctrl}^{\Gamma}_n(f_1,\ldots,f_n):
       \mathbf R_n\otimes(\Gamma\mathbin{\cdot}D)
       \to\mathbf R_n\otimes P},
 \qquad n\geq1,\quad D,P\ \text{positive}.
 \label{eq:QC-control-formation}
\end{equation}
Put \(\operatorname{ctrl}_n:=\operatorname{ctrl}^{\varnothing}_n\).

Let \(\equiv\) be the least typed congruence, closed under every term
former, that contains the symmetric \(*\)-autonomous equations, the
register equations, and the following relations.  For
\[
 \begin{gathered}
 f,f_i:\Gamma\mathbin{\cdot}D\to P,
 \qquad g_i:P\to R,\\
 h:\Gamma'\mathbin{\cdot}D'\to\Gamma\mathbin{\cdot}D,
 \qquad D,D',P,R\ \text{positive},
 \end{gathered}
\]
impose
\begin{align}
 \operatorname{ctrl}^{\Gamma}_n(f,\ldots,f)
 &\equiv\id_{\mathbf R_n}\otimes f,
 \label{eq:QC-control-uniform}\\
 \operatorname{ctrl}_n(g_1,\ldots,g_n)
 \circ\operatorname{ctrl}^{\Gamma}_n(f_1,\ldots,f_n)
 &\equiv\operatorname{ctrl}^{\Gamma}_n
 (g_1\circ f_1,\ldots,g_n\circ f_n),
 \label{eq:QC-control-composition}\\
 \operatorname{ctrl}^{\Gamma}_n(f_1,\ldots,f_n)
 \circ(\id_{\mathbf R_n}\otimes h)
 &\equiv\operatorname{ctrl}^{\Gamma'}_n
 (f_1\circ h,\ldots,f_n\circ h).
 \label{eq:QC-control-prenaturality}
\end{align}
For \(\rho\in S_n\), let
\(\Pi_\rho:\CC[n]\to\CC[n]\) be the permutation unitary defined by
\(\Pi_\rho\lvert i\rangle=\lvert\rho(i)\rangle\), and impose
\begin{equation}
 (u_{\Pi_\rho}\otimes\id_P)
 \circ\operatorname{ctrl}^{\Gamma}_n(f_1,\ldots,f_n)
 \circ
 (u_{\Pi_{\rho^{-1}}}\otimes
  \id_{\Gamma\mathbin{\cdot}D})
 \equiv
 \operatorname{ctrl}^{\Gamma}_n
 (f_{\rho^{-1}(1)},\ldots,f_{\rho^{-1}(n)}).
 \label{eq:QC-control-permutation}
\end{equation}
For shape-compatible positive \(D,P\) and unitaries
\(U_i:\CC[I_D]\to\CC[I_P]\), also impose
\begin{equation}
 \operatorname{ctrl}_n
 (u_{U_1}^{D,P},\ldots,u_{U_n}^{D,P})
 \equiv
 u_{\bigoplus_iU_i}^{\mathbf R_n\otimes D,\,
                         \mathbf R_n\otimes P},
 \label{eq:QC-control-register-overlap}
\end{equation}
where the block sum uses the fixed lexicographic order on
\([n]\times I_D\) and \([n]\times I_P\).
\end{definition}
The last relation fixes the intended overlap between contextual
control and compound-register gates.  In particular, uniformity gives
\[
 \id_{\mathbf R_n}\otimes u_U^{D,P}
 \equiv
 u_{\bigoplus_iU}^{\mathbf R_n\otimes D,\,
                      \mathbf R_n\otimes P};
\]
this is selector--tensor compatibility, not arbitrary tensoriality of
the gate presentation.  The positive-codomain restriction keeps the
controlled output on a pure axis.

\begin{definition}[The coherent quantum core \(\QC\)]
\label{def:basic-QC}
\label{def:signed-QC}
\label{def:QC-doctrine-equations}
The objects of \(\QC\) are the raw types.  A term is \emph{raw} when
every formula in its derivation, including
every control premise and composition intermediate, is a raw type.
Writing \(\mathsf{RawTerm}(A,B)\) for these terms, for raw types
\(A,B\), \(\QC(A,B)\) is the set of \(\equiv\)-equivalence
classes in \(\mathsf{Term}(A,B)\) that contain a raw representative:
\[
 \QC(A,B)
 :=
 \bigl\{
 [t]_{\equiv}
 \bigm|
 t\in\mathsf{Term}(A,B),\qquad
 \exists r\in\mathsf{RawTerm}(A,B),\ r\equiv t
 \bigr\}.
\]
\end{definition}
Thus \(\QC\) is the image of the raw term calculus under the single
global ambient congruence, rather than all equivalence classes between
raw objects.  Raw
representatives are closed under identities, composition, tensor,
\(\parr\), and duality.  The standard unit-free derivations of
\[
 \eta_{X,A}:X\to A^*\parr(X\otimes A),
 \qquad
 \ev_{A,B}:(A^*\parr B)\otimes A\to B
\]
show that currying and uncurrying also restrict to \(\QC\).  We call
this inherited multiplicative, dual, and closed structure
unit-free \(*\)-autonomous.\footnote{Removing the units from a \(*\)-autonomous presentation
is subtler than deleting two objects~\cite{HoustonHughesSchalk2005};
the full version develops the unit-free closed structure.}
The full version gives the finitary presentation and the unit-free
closed maps.

For fixed \(n\), Equations~\eqref{eq:QC-control-uniform}
and~\eqref{eq:QC-control-composition} make
\(P\mapsto\mathbf R_n\otimes P\),
\((f_i)_{i\in[n]}\mapsto\operatorname{ctrl}_n(f_1,\ldots,f_n)\)
functorial on \(n\)-tuples of parallel morphisms between positive
types, and Equation~\eqref{eq:QC-control-permutation} is its selector
covariance: permuting the branches by \(\rho\) is conjugation by
\(u_{\Pi_\rho}\otimes\id\).
Equation~\eqref{eq:QC-control-prenaturality} extends this block
action to shared inputs \(\Gamma\mathbin{\cdot}D\). Since control
retains one selector port whereas pointwise tensor produces two, the
assignment is not tensor-preserving.

Put \(Q:=(1,0)\). Since \(\mathbf H_n=Q^{\oplus n}\), the carrier
normal form supplies, for every raw \(A\), the selector-row display
\[
 d_{n,A}:\mathbf H_n\otimes O(A)
 \xrightarrow{\cong}
 \bigoplus_{i=1}^n\bigl(Q\otimes O(A)\bigr).
\]
For positive \(P\), let \(\phi_{n,P}\) be the canonical
coefficient-one lexicographic flattening.  The recursive row convention
gives
\begin{equation}
 d_{\mathbf R_n\otimes P}
 =
 \phi_{n,P}
 \left(\bigoplus_{s=1}^n(\id_Q\otimes d_P)\right)
 d_{n,P}.
 \label{eq:QC-selector-display-compatibility}
\end{equation}

Regard \(\Perm(\CC)\) as \(*\)-autonomous via its dagger compact
structure.

\begin{definition}[Carrier assignment]
\label{def:QC-semantics}
Extend \(O\) structurally to the ambient objects and interpret the
structural arrows canonically. For an ambient term
\(f\in\mathsf{Term}(X,Y)\), write
\(\widetilde{\llbracket f\rrbracket}:O(X)\to O(Y)\) for its carrier.
For shape-compatible positive \(P,Q\) and unitary
\(U:\CC[I_P]\to\CC[I_Q]\), assign the register gates by
\begin{equation}
 d_Q\,\widetilde{\llbracket u_U^{P,Q}\rrbracket}\,d_P^{-1}
 =
 \bigl(U_{\mathbf j,\mathbf i}\,
       \id_{(k_P,0)}\bigr)_{
       \mathbf j\in I_Q,\,\mathbf i\in I_P}.
 \label{eq:QC-register-unitary-carrier}
\end{equation}
Negative gates receive the dual carrier maps.  For
\(f_i:\Gamma\mathbin{\cdot}D\to P\), set
\begin{equation}
 \widetilde{\llbracket
   \operatorname{ctrl}^{\Gamma}_n(f_1,\ldots,f_n)
 \rrbracket}
 =
 d_{n,P}^{-1}
 \left(
  \bigoplus_{i=1}^n
  \bigl(\id_Q\otimes
        \widetilde{\llbracket f_i\rrbracket}\bigr)
 \right)
 d_{n,\Gamma\mathbin{\cdot}D}.
 \label{eq:QC-control-carrier}
\end{equation}
\end{definition}

Formula~\eqref{eq:QC-control-carrier} keeps the qudit port in every
branch; only its address selects a branch. It is therefore not the
arbitrary carrier direct sum excluded by the first counterexample of
Section~\ref{subsec:QC-why-coherence}.

\begin{theorem}[Carrier functor]
\label{thm:basic-structure}
The carrier assignments on ambient terms respect \(\equiv\) and
therefore induce uniquely a functor
\(\llbracket-\rrbracket:\QC\longrightarrow\Perm(\CC)\).
This functor preserves the inherited symmetric multiplicative, dual,
and closed structure.
\end{theorem}

\begin{proof}
Structural recursion defines the carrier on finite terms.  In
selector-row coordinates, uniformity, composition, prenaturality, and
permutation covariance are the corresponding block-diagonal
identities.  Equation~\eqref{eq:QC-selector-display-compatibility}
makes the register overlap the block matrix
\(\bigoplus_iU_i\).  Hence the assignment respects every generating
relation and descends uniquely through \(\equiv\).
\extendedonly{The coordinate verification is given in
Appendix~\ref{prop:QC-carrier-descent}.}
\end{proof}

For \(f=[t]_{\equiv}\), set
\[
 \llbracket f\rrbracket
 :=\widetilde{\llbracket t\rrbracket}.
\]
This is well defined by Theorem~\ref{thm:basic-structure}.

\subsection{An Applied Contextual Quantum Switch}
\label{subsec:contextual-switch-v2}

Fix a positive raw type \(D\), and put \(E:=D^*\parr D\). Define
\[
 c:=
 \ev_{D,D}\circ(\id_E\otimes\ev_{D,D}):
 E\otimes E\otimes D\longrightarrow D,
\]
and set
\[
 c_0:=c\circ(\sigma_{E,E}\otimes\id_D),
 \qquad
 c_1:=c.
\]
Thus \(c_0(f,g,x)=g(fx)\) and \(c_1(f,g,x)=f(gx)\). The applied
switch is
\begin{equation}
 \mathsf{QSwitch}^{\eta}_{D}
 :=\operatorname{ctrl}^{(E,E)}_2(c_0,c_1):
 \mathbf R_2\otimes E\otimes E\otimes D
 \longrightarrow
 \mathbf R_2\otimes D.
 \label{eq:QC-qswitch}
\end{equation}
Its curried form has type
\[
 \mathsf{QSwitch}_D:
 E\otimes E
 \longrightarrow
 \mathbf R_2^*\parr
 \bigl(D^*\parr(\mathbf R_2\otimes D)\bigr).
\]
This realizes, within the present essentially-unitary fragment, the
\(\mathtt{qcase}\) quantum switch of
Barsse, P\'echoux, and Perdrix~\cite{BarssePechouxPerdrix2026},
with an arbitrary positive payload \(D\). In selector-row coordinates,
the two diagonal blocks are the two causal orders; a unitary on
\(\mathbf R_2\) coherently superposes them.

Thus the positive-output discipline still realizes coherent control of
causal order: the shared higher-order arguments occur in both branches,
while the selector is retained in the codomain
\(\mathbf R_2\otimes D\).

\subsection{Essential Unitarity of the Coherent Syntax}
\label{sec:calculus}

Section~\ref{sec:eu-predicate} assigns a carrier matrix to every
nonempty qudit--MLL boundary and a full numerical matrix to every
balanced one.  A source morphism is essentially unitary when its full
matrix is dagger-unitary.  This subsection states the main theorem;
the complete proofs appear in the full version of this paper.

Throughout, \(I:=\Unit\) denotes the tensor unit of \(\Perm(\CC)\).
For a source sequent \(\Sigma=(X_1,\ldots,X_s)\), put
\(O(\Sigma):=O(X_1)\otimes\cdots\otimes O(X_s)\).  In every boundary
datum imported from Definition~\ref{def:qudit-mll-sequent-matrix},
the subscript \(\Sigma\) abbreviates the carrier sequent
\((O(X_1),\ldots,O(X_s))\), and for a source type \(X\) we write
\(\mathsf A_X^\pm:=\mathsf A_{O(X)}^\pm\).  For \(h:A\to B\), let
\[
 a_h:=
 \operatorname{name}_{\Perm(\CC)}(\llbracket h\rrbracket):
 I\longrightarrow O(A^*,B),
 \qquad
 M(h):=\mathcal Q_{(A^*,B)}(a_h).
\]
Thus
\[
 M(h):
 \mathsf A_A^+\otimes\mathsf A_B^-
 \longrightarrow
 \mathsf A_A^-\otimes\mathsf A_B^+.
\]
If \(A\) and \(B\) are positive, unit coherence identifies \(M(h)\)
with \(\llbracket h\rrbracket\); if \(A=P^*\) and \(B=Q^*\) with
\(P,Q\) positive, it identifies \(M(h)\) with the compact transpose
\(\llbracket h\rrbracket^*:O(Q)\to O(P)\).
For every \(h\) in \(\QC\), the positive and negative widths of this
boundary agree and are at least \(1\), so the fixed-width numerical
realization of Definition~\ref{def:qudit-mll-sequent-matrix} applies.
Write \(\mathbf M(h):=\mathbf Q_{(A^*,B)}(a_h)\) for the full
qudit--MLL matrix.  By definition, \(\operatorname{EU}(h)\) means
that \(\mathbf M(h)\) is dagger-unitary.

\begin{theorem}[Essential unitarity of \(\QC\)]
\label{thm:calculus-EU}
For every \(h:A\to B\) in \(\QC\), the polarized carrier matrix
\(M(h)\) depends only on \(h\) and satisfies
\[
 M(h)^\dagger M(h)
 =\id_{\mathsf A_A^+\otimes\mathsf A_B^-},
 \qquad
 M(h)M(h)^\dagger
 =\id_{\mathsf A_A^-\otimes\mathsf A_B^+}.
\]
Consequently \(\mathbf M(h)\) is dagger-unitary, so
\(\operatorname{EU}(h)\).
\end{theorem}

The proof has four steps.  A source expression is translated to an
administrative expression with the same compact-name carrier point.
A categorical normalization replaces its compositions at mixed
intermediate types by carrier-equal compositions through pure,
meaning one-polarity, intermediate types.  On pure interfaces, the
qudit--MLL matrix preserves composition and dagger.  Finally,
multiplicative base maps, positive-register gates, and contextual
control contribute, respectively, coefficient-one matching matrices,
unitary qudit matrices, and synchronized block-diagonal matrices.

\section{One-Slot Supermaps}
\label{sec:expr-v2}\label{sec:expressiveness}

A one-slot context uses its argument once along every selected control
branch. Composition and tensor place fixed programs around that slot;
contextual control chooses among a synchronized family without copying
its input.

Fix a slot \(\Box:A\to B\) and a class
\(\mathsf{Const}\subseteq\operatorname{Mor}(\QC)\) of well-typed constant
programs. Resource-linear contexts are generated from \(\Box\) by
composition and tensor with constants \(K\in\mathsf{Const}\) on
either side, and by
\(\operatorname{ctrl}^{\Gamma}_n(D_1[\Box],\ldots,D_n[\Box])\),
with every expression well typed. In the control clause, \(E=\Gamma\mathbin{\cdot}D\) for a positive \(D\); the
contexts \(D_i[\Box]:E\to P\) have a common positive
codomain \(P\), and the resulting context has type
\(\mathbf R_n\otimes E\to\mathbf R_n\otimes P\).
The grammar is closed under the typed currying and uncurrying
bijections of \(\QC\), each preserving its single slot.

Every well-typed resource-linear context
\(C[\Box]:E\to F\), with \(\Box:A\to B\), defines
\(\Phi_C:\QC(A,B)\to\QC(E,F)\), \(\Phi_C(f):=C[f]\), and admits a
\(\CC\)-linear carrier map
\[
 \widetilde\Phi_C:
 \Perm(\CC)\bigl(O(A),O(B)\bigr)
 \longrightarrow
 \Perm(\CC)\bigl(O(E),O(F)\bigr)
\]
satisfying
\(\widetilde\Phi_C(\llbracket f\rrbracket)
=\llbracket\Phi_C(f)\rrbracket\) for every \(f:A\to B\) in \(\QC\);
every \(\Phi_C(f)\) is essentially unitary.

\begin{proof}
Induct on the context. The slot gives \(\widetilde\Phi_\Box=\id\).
Writing \(k:=\llbracket K\rrbracket\), the four multiplicative clauses
send \(x\) to \(\widetilde\Phi_D(x)\circ k\),
\(k\circ\widetilde\Phi_D(x)\),
\(\widetilde\Phi_D(x)\otimes k\), and
\(k\otimes\widetilde\Phi_D(x)\), respectively.
For a control clause with \(D_i[\Box]:E\to P\), set
\[
 \widetilde\Phi_C(x)
 :=d_{n,P}^{-1}
 \left(
  \bigoplus_{i=1}^n
  \bigl(\id_Q\otimes\widetilde\Phi_{D_i}(x)\bigr)
 \right)d_{n,E}.
\]
These maps are linear by bilinearity and finite biproduct assembly;
currying and uncurrying use the linear compact name and unname
isomorphisms. The carrier equations give the substitution identity.
Closure of \(\QC\) and Theorem~\ref{thm:calculus-EU} give the remaining
claims.
\end{proof}

The source calculus realizes the two unitary stages of an exposed-memory
one-slot dilation. Memory initialization and partial trace remain in
the external Hilbert-space/channel semantics.

For a register \(X=\mathbf R_d\), write \(H_X\cong\CC^d\) for
its external Hilbert space. Channels are completely positive
trace-preserving maps
\(\mathcal B(H_{\mathrm{in}})\to\mathcal B(H_{\mathrm{out}})\),
and a deterministic one-slot supermap is a completely CP-preserving
linear map sending channels to
channels~\cite{Chiribella2008Supermaps}.

\begin{theorem}[Internal realization of equal-ratio one-slot supermaps]
\label{thm:equal-ratio-supermaps-v2}
\label{thm:equal-dim-supermaps}
Let \(A_i=\mathbf R_{a_i}\) and \(B_i=\mathbf R_{b_i}\) for
\(i=0,1\), with \(a_0b_1=a_1b_0\), and let \(\Phi\) be a deterministic one-slot
supermap from channels
\(\mathcal B(H_{A_0})\to\mathcal B(H_{A_1})\) to channels
\(\mathcal B(H_{B_0})\to\mathcal B(H_{B_1})\). Then, for some
registers \(M=\mathbf R_m\) and \(N=\mathbf R_n\), there are morphisms
in \(\QC\)
\[
 V_1^{\QC}:B_0\otimes M\longrightarrow A_0\otimes N,
 \qquad
 V_2^{\QC}:A_1\otimes N\longrightarrow B_1\otimes M
\]
whose carriers are unitary extensions of the two isometries in a
Chiribella realization of \(\Phi\). For every \(f:A_0\to A_1\) in
\(\QC\),
\[
 \Phi^\sharp(f):=
 V_2^{\QC}\circ(f\otimes\id_N)\circ V_1^{\QC}:
 B_0\otimes M\longrightarrow B_1\otimes M
\]
is a morphism of \(\QC\) and is essentially unitary.
\end{theorem}

\begin{proof}
By \cite[Theorem~1, Eq.~(23)]{Chiribella2008Supermaps}, choose
isometries
\[
 V:H_{B_0}\longrightarrow H_{A_0}\otimes H_{N_0},
 \qquad
 W:H_{A_1}\otimes H_{N_0}
   \longrightarrow H_{B_1}\otimes H_{M_0}
\]
representing \(\Phi\).
Write
\(b_0/a_0=b_1/a_1=r/s\) in lowest terms. Choose \(L\) so that
\(rL\geq\dim H_{N_0}\) and \(sL\geq\dim H_{M_0}\), and put
\(n:=rL\), \(m:=sL\). Then \(b_0m=a_0n\) and \(a_1n=b_1m\).
Choose isometric embeddings \(j_M:H_{M_0}\hookrightarrow H_M\),
\(j_N:H_{N_0}\hookrightarrow H_N\), and a unit vector
\(\ket{0}_M\in H_M\). The
isometries
\[
 \psi\otimes\ket{0}_M
 \longmapsto(\id_{H_{A_0}}\otimes j_N)V\psi,
 \qquad
 (\id_{H_{A_1}}\otimes j_N)\xi
 \longmapsto(\id_{H_{B_1}}\otimes j_M)W\xi
\]
extend to unitaries \(\widetilde V_1\) and \(\widetilde V_2\) on the
equal-dimensional ambient spaces.

Both source--target pairs are positive, have port width \(2\), and
have equal address cardinality. If \(G_1,G_2\) are the matrices of
\(\widetilde V_1,\widetilde V_2\) in the lexicographic qudit bases,
Definition~\ref{def:QC-register-unitaries} supplies
\[
 V_1^{\QC}:=u_{G_1}^{B_0\otimes M,\,A_0\otimes N},
 \qquad
 V_2^{\QC}:=u_{G_2}^{A_1\otimes N,\,B_1\otimes M}.
\]
Their carriers are the required unitary extensions. Closure of \(\QC\)
and Theorem~\ref{thm:calculus-EU} give the conclusion.
\end{proof}

The stage construction does not require equal endpoint dimensions. By
Theorem~\ref{thm:calculus-EU}, however, \(\QC(A_0,A_1)\ne\varnothing\)
forces \(a_0=a_1\); the equal-ratio condition then forces \(b_0=b_1\).

For \(g:P\to Q\) in \(\QC\) between positive types, boundary balance
gives \(k_P=k_Q=:k\). Define its full fixed-width matrix by
\[
 \widehat U_g
 :=\mathsf B_k\!\left(
 d_Q\,\llbracket g\rrbracket\,d_P^{-1}
 \right):
 \CC[I_P]\otimes\CC[k]\longrightarrow
 \CC[I_Q]\otimes\CC[k].
\]
Proposition~\ref{prop:first-order-representation} gives
\(\widehat U_{h\circ g}=\widehat U_h\widehat U_g\), and
Theorem~\ref{thm:calculus-EU} makes \(\widehat U_g\) unitary. For
\(k=1\), identify \(\widehat U_g\) with \(U_g:\CC[I_P]\to\CC[I_Q]\).
For a unitary \(U\), write
\(\operatorname{Ad}_U(\rho):=U\rho U^\dagger\).

\begin{corollary}[External readout of the internal lift]
\label{cor:equal-ratio-supermap-readout}
For the realization of Theorem~\ref{thm:equal-ratio-supermaps-v2}
and every \(f:A_0\to A_1\) in \(\QC\), the full matrix factors as
\(\widehat U_{\Phi^\sharp(f)}=U_{\Phi^\sharp(f)}\otimes\id_{\CC[2]}\)
with
\(U_{\Phi^\sharp(f)}=\widetilde V_2(U_f\otimes\id_{H_N})\widetilde V_1\).
There is a pure state \(\omega_M\) on \(H_M\), independent of \(f\),
such that, for every such \(f\) and every density operator \(\rho\) on
\(H_{B_0}\),
\(\Phi(\operatorname{Ad}_{U_f})(\rho)
 =\operatorname{Tr}_{H_M}\!\bigl[
 \operatorname{Ad}_{U_{\Phi^\sharp(f)}}(\rho\otimes\omega_M)
 \bigr]\).
\end{corollary}

\begin{proof}
Use the embeddings and unitary extensions constructed in the proof of
Theorem~\ref{thm:equal-ratio-supermaps-v2}, and put
\(\omega_M:=\ket{0}_M\!\bra{0}_M\). For every channel
\(\mathcal F:\mathcal B(H_{A_0})\to\mathcal B(H_{A_1})\), the
extension identities give
\[
 \Phi(\mathcal F)(\rho)
 =\operatorname{Tr}_{H_M}\!\left[
 \widetilde V_2
 (\mathcal F\otimes\operatorname{id}_{\mathcal B(H_N)})
 \bigl(
  \widetilde V_1(\rho\otimes\omega_M)\widetilde V_1^\dagger
 \bigr)
 \widetilde V_2^\dagger
 \right].
\]
Here \(\mathcal F\otimes\operatorname{id}\) preserves the
\(j_N\)-supported block; the defining extension identity for
\(\widetilde V_2\) then produces the \(j_M\)-embedded sequential output,
and \(\operatorname{Tr}_{H_M}\) removes this final isometric embedding
\cite[Eq.~(23)]{Chiribella2008Supermaps}. Taking
\(\mathcal F=\operatorname{Ad}_{U_f}\), the register-gate carrier
equation, monoidality, and
Proposition~\ref{prop:first-order-representation} give the displayed
factorization of \(\widehat U_{\Phi^\sharp(f)}\), hence the result.
\end{proof}

The argument isolates the source discipline that makes essential
unitarity possible: multiplicative structure, positive-register
gates, and contextual qudit control reduce to dagger-unitary matrices
composed through pure interfaces. An internal treatment of
measurement, partial trace, and mixed-state semantics is left to
future work.

\bibliographystyle{eptcs}
\bibliography{merged}

\begin{thebibliography}{10}
\providecommand{\bibitemdeclare}[2]{}
\providecommand{\surnamestart}{}
\providecommand{\surnameend}{}
\providecommand{\urlprefix}{Available at }
\providecommand{\url}[1]{\texttt{#1}}
\providecommand{\href}[2]{\texttt{#2}}
\providecommand{\urlalt}[2]{\href{#1}{#2}}
\providecommand{\doi}[1]{doi:\urlalt{https://doi.org/#1}{#1}}
\providecommand{\eprint}[1]{arXiv:\urlalt{https://arxiv.org/abs/#1}{#1}}
\providecommand{\bibinfo}[2]{#2}

\bibitemdeclare{inproceedings}{AbramskyAS}
\bibitem{AbramskyAS}
\bibinfo{author}{Samson \surnamestart Abramsky\surnameend}
  (\bibinfo{year}{2005}): \emph{\bibinfo{title}{Abstract scalars, loops, and
  free traced and strongly compact closed categories}}.
\newblock In: {\slshape \bibinfo{booktitle}{Proceedings of the First
  International Conference on Algebra and Coalgebra in Computer Science}},
  \bibinfo{series}{CALCO'05}, \bibinfo{publisher}{Springer-Verlag},
  \bibinfo{address}{Berlin, Heidelberg}, p. \bibinfo{pages}{1–29},
  \doi{10.1007/11548133_1}.

\bibitemdeclare{inproceedings}{abramskyCoecke2004}
\bibitem{abramskyCoecke2004}
\bibinfo{author}{Samson \surnamestart Abramsky\surnameend} \&
  \bibinfo{author}{Bob \surnamestart Coecke\surnameend} (\bibinfo{year}{2004}):
  \emph{\bibinfo{title}{A categorical semantics of quantum protocols}}.
\newblock In: {\slshape \bibinfo{booktitle}{Proceedings of the 19th Annual IEEE
  Symposium on Logic in Computer Science (LICS '04)}}, \bibinfo{publisher}{IEEE
  Computer Society}, pp. \bibinfo{pages}{415--425},
  \doi{10.1109/LICS.2004.1319636}.

\bibitemdeclare{article}{A2014}
\bibitem{A2014}
\bibinfo{author}{Mateus \surnamestart Ara\'{u}jo\surnameend},
  \bibinfo{author}{Fabio \surnamestart Costa\surnameend} \&
  \bibinfo{author}{\v{C}aslav \surnamestart Brukner\surnameend}
  (\bibinfo{year}{2014}): \emph{\bibinfo{title}{Computational Advantage from
  Quantum-Controlled Ordering of Gates}}.
\newblock {\slshape \bibinfo{journal}{Physical Review Letters}}
  \bibinfo{volume}{113}(\bibinfo{number}{25}), p. \bibinfo{pages}{250402},
  \doi{10.1103/PhysRevLett.113.250402}.
\newblock \eprint{1401.8127}.

\bibitemdeclare{misc}{BarssePechouxPerdrix2026}
\bibitem{BarssePechouxPerdrix2026}
\bibinfo{author}{Kathleen \surnamestart Barsse\surnameend},
  \bibinfo{author}{Romain \surnamestart P{\'e}choux\surnameend} \&
  \bibinfo{author}{Simon \surnamestart Perdrix\surnameend}
  (\bibinfo{year}{2026}): \emph{\bibinfo{title}{Higher-Order Programs with
  Indefinite Causal Orders: A Linear Approach to Coherent Control of Quantum
  Processes}}, \doi{10.48550/arXiv.2607.09534}.
\newblock \eprint{2607.09534}.

\bibitemdeclare{article}{BCST1996}
\bibitem{BCST1996}
\bibinfo{author}{R.~F. \surnamestart Blute\surnameend},
  \bibinfo{author}{J.~R.~B. \surnamestart Cockett\surnameend},
  \bibinfo{author}{R.~A.~G. \surnamestart Seely\surnameend} \&
  \bibinfo{author}{T.~H. \surnamestart Trimble\surnameend}
  (\bibinfo{year}{1996}): \emph{\bibinfo{title}{Natural deduction and coherence
  for weakly distributive categories}}.
\newblock {\slshape \bibinfo{journal}{Journal of Pure and Applied Algebra}}
  \bibinfo{volume}{113}(\bibinfo{number}{3}), pp. \bibinfo{pages}{229--296},
  \doi{10.1016/0022-4049(95)00159-X}.

\bibitemdeclare{article}{Chiribella2008}
\bibitem{Chiribella2008}
\bibinfo{author}{Giulio \surnamestart Chiribella\surnameend},
  \bibinfo{author}{Giacomo~Mauro \surnamestart D'Ariano\surnameend} \&
  \bibinfo{author}{Paolo \surnamestart Perinotti\surnameend}
  (\bibinfo{year}{2008}): \emph{\bibinfo{title}{Quantum Circuit Architecture}}.
\newblock {\slshape \bibinfo{journal}{Phys. Rev. Lett.}} \bibinfo{volume}{101},
  p. \bibinfo{pages}{060401}, \doi{10.1103/PhysRevLett.101.060401}.

\bibitemdeclare{article}{Chiribella2008Supermaps}
\bibitem{Chiribella2008Supermaps}
\bibinfo{author}{Giulio \surnamestart Chiribella\surnameend},
  \bibinfo{author}{Giacomo~Mauro \surnamestart D'Ariano\surnameend} \&
  \bibinfo{author}{Paolo \surnamestart Perinotti\surnameend}
  (\bibinfo{year}{2008}): \emph{\bibinfo{title}{Transforming quantum
  operations: Quantum supermaps}}.
\newblock {\slshape \bibinfo{journal}{Europhysics Letters}}
  \bibinfo{volume}{83}(\bibinfo{number}{3}), p. \bibinfo{pages}{30004},
  \doi{10.1209/0295-5075/83/30004}.

\bibitemdeclare{article}{Chiribella2009}
\bibitem{Chiribella2009}
\bibinfo{author}{Giulio \surnamestart Chiribella\surnameend},
  \bibinfo{author}{Giacomo~Mauro \surnamestart D'Ariano\surnameend} \&
  \bibinfo{author}{Paolo \surnamestart Perinotti\surnameend}
  (\bibinfo{year}{2009}): \emph{\bibinfo{title}{Theoretical framework for
  quantum networks}}.
\newblock {\slshape \bibinfo{journal}{Phys. Rev. A}} \bibinfo{volume}{80}, p.
  \bibinfo{pages}{022339}, \doi{10.1103/PhysRevA.80.022339}.

\bibitemdeclare{article}{chiribella2013quantum}
\bibitem{chiribella2013quantum}
\bibinfo{author}{Giulio \surnamestart Chiribella\surnameend},
  \bibinfo{author}{Giacomo~Mauro \surnamestart D'Ariano\surnameend},
  \bibinfo{author}{Paolo \surnamestart Perinotti\surnameend} \&
  \bibinfo{author}{Benoit \surnamestart Valiron\surnameend}
  (\bibinfo{year}{2013}): \emph{\bibinfo{title}{Quantum computations without
  definite causal structure}}.
\newblock {\slshape \bibinfo{journal}{Phys. Rev. A}}
  \bibinfo{volume}{88}(\bibinfo{number}{2}), p. \bibinfo{pages}{022318},
  \doi{10.1103/PhysRevA.88.022318}.

\bibitemdeclare{article}{CockettSeely1997WDC}
\bibitem{CockettSeely1997WDC}
\bibinfo{author}{J.~R.~B. \surnamestart Cockett\surnameend} \&
  \bibinfo{author}{R.~A.~G. \surnamestart Seely\surnameend}
  (\bibinfo{year}{1997}): \emph{\bibinfo{title}{Weakly distributive
  categories}}.
\newblock {\slshape \bibinfo{journal}{Journal of Pure and Applied Algebra}}
  \bibinfo{volume}{114}(\bibinfo{number}{2}), pp. \bibinfo{pages}{133--173},
  \doi{10.1016/0022-4049(95)00160-3}.

\bibitemdeclare{inproceedings}{ZX}
\bibitem{ZX}
\bibinfo{author}{Bob \surnamestart Coecke\surnameend} \& \bibinfo{author}{Ross
  \surnamestart Duncan\surnameend} (\bibinfo{year}{2008}):
  \emph{\bibinfo{title}{Interacting Quantum Observables}}.
\newblock In \bibinfo{editor}{Luca \surnamestart Aceto\surnameend},
  \bibinfo{editor}{Ivan \surnamestart Damg{\aa}rd\surnameend},
  \bibinfo{editor}{Leslie~Ann \surnamestart Goldberg\surnameend},
  \bibinfo{editor}{Magn{\'u}s~M. \surnamestart Halld{\'o}rsson\surnameend},
  \bibinfo{editor}{Anna \surnamestart Ing{\'o}lfsd{\'o}ttir\surnameend} \&
  \bibinfo{editor}{Igor \surnamestart Walukiewicz\surnameend}, editors:
  {\slshape \bibinfo{booktitle}{Automata, Languages and Programming}},
  \bibinfo{publisher}{Springer Berlin Heidelberg}, \bibinfo{address}{Berlin,
  Heidelberg}, pp. \bibinfo{pages}{298--310},
  \doi{10.1007/978-3-540-70583-3_25}.

\bibitemdeclare{book}{CoeckeKissinger2017}
\bibitem{CoeckeKissinger2017}
\bibinfo{author}{Bob \surnamestart Coecke\surnameend} \& \bibinfo{author}{Aleks
  \surnamestart Kissinger\surnameend} (\bibinfo{year}{2017}):
  \emph{\bibinfo{title}{Picturing Quantum Processes: A First Course in Quantum
  Theory and Diagrammatic Reasoning}}.
\newblock \bibinfo{publisher}{Cambridge University Press},
  \doi{10.1017/9781316219317}.

\bibitemdeclare{article}{Coecke2013}
\bibitem{Coecke2013}
\bibinfo{author}{Bob \surnamestart Coecke\surnameend}, \bibinfo{author}{Dusko
  \surnamestart Pavlovic\surnameend} \& \bibinfo{author}{Jamie \surnamestart
  Vicary\surnameend} (\bibinfo{year}{2013}): \emph{\bibinfo{title}{{A new
  description of orthogonal bases}}}.
\newblock {\slshape \bibinfo{journal}{Mathematical Structures in Computer
  Science}} \bibinfo{volume}{23}(\bibinfo{number}{3}), pp.
  \bibinfo{pages}{555--567}, \doi{10.1017/S0960129512000047}.
\newblock \eprint{0810.0812}.

\bibitemdeclare{inproceedings}{HeffordWilson2024}
\bibitem{HeffordWilson2024}
\bibinfo{author}{James \surnamestart Hefford\surnameend} \&
  \bibinfo{author}{Matt \surnamestart Wilson\surnameend}
  (\bibinfo{year}{2024}): \emph{\bibinfo{title}{A Profunctorial Semantics for
  Quantum Supermaps}}.
\newblock In: {\slshape \bibinfo{booktitle}{Proceedings of the 39th Annual
  ACM/IEEE Symposium on Logic in Computer Science (LICS '24)}},
  \bibinfo{publisher}{ACM}, pp. \bibinfo{pages}{43:1--43:15},
  \doi{10.1145/3661814.3662123}.
\newblock \eprint{2402.02997}.

\bibitemdeclare{inproceedings}{HirataTsukada2026}
\bibitem{HirataTsukada2026}
\bibinfo{author}{Kengo \surnamestart Hirata\surnameend} \&
  \bibinfo{author}{Takeshi \surnamestart Tsukada\surnameend}
  (\bibinfo{year}{2026}): \emph{\bibinfo{title}{Causality in Pure Quantum
  Computation with Quantum Control}}.
\newblock In: {\slshape \bibinfo{booktitle}{41st Annual Symposium on Logic in
  Computer Science (LICS 2026)}}, {\slshape \bibinfo{series}{Leibniz
  International Proceedings in Informatics (LIPIcs)}} \bibinfo{volume}{380},
  \bibinfo{publisher}{Schloss Dagstuhl--Leibniz-Zentrum f{\"u}r Informatik},
  pp. \bibinfo{pages}{57:1--57:28}, \doi{10.4230/LIPIcs.LICS.2026.57}.

\bibitemdeclare{misc}{HoustonHughesSchalk2005}
\bibitem{HoustonHughesSchalk2005}
\bibinfo{author}{Robin \surnamestart Houston\surnameend},
  \bibinfo{author}{Dominic J.~D. \surnamestart Hughes\surnameend} \&
  \bibinfo{author}{Andrea \surnamestart Schalk\surnameend}
  (\bibinfo{year}{2005}): \emph{\bibinfo{title}{Modelling linear logic without
  units (preliminary results)}}.
\newblock \eprint{math/0504037}.

\bibitemdeclare{article}{Hughes2012}
\bibitem{Hughes2012}
\bibinfo{author}{Dominic J.~D. \surnamestart Hughes\surnameend}
  (\bibinfo{year}{2012}): \emph{\bibinfo{title}{Simple free star-autonomous
  categories and full coherence}}.
\newblock {\slshape \bibinfo{journal}{Journal of Pure and Applied Algebra}}
  \bibinfo{volume}{216}(\bibinfo{number}{11}), pp. \bibinfo{pages}{2386--2410},
  \doi{10.1016/j.jpaa.2012.03.020}.

\bibitemdeclare{article}{coherence}
\bibitem{coherence}
\bibinfo{author}{G.M. \surnamestart Kelly\surnameend} \& \bibinfo{author}{M.L.
  \surnamestart Laplaza\surnameend} (\bibinfo{year}{1980}):
  \emph{\bibinfo{title}{Coherence for compact closed categories}}.
\newblock {\slshape \bibinfo{journal}{Journal of Pure and Applied Algebra}}
  \bibinfo{volume}{19}, pp. \bibinfo{pages}{193--213},
  \doi{10.1016/0022-4049(80)90101-2}.

\bibitemdeclare{article}{KissingerUijlen2019}
\bibitem{KissingerUijlen2019}
\bibinfo{author}{Aleks \surnamestart Kissinger\surnameend} \&
  \bibinfo{author}{Sander \surnamestart Uijlen\surnameend}
  (\bibinfo{year}{2019}): \emph{\bibinfo{title}{A categorical semantics for
  causal structure}}.
\newblock {\slshape \bibinfo{journal}{Logical Methods in Computer Science}}
  \bibinfo{volume}{15}(\bibinfo{number}{3}), \doi{10.23638/LMCS-15(3:15)2019}.
\newblock \eprint{1701.04732}.

\bibitemdeclare{incollection}{Laplaza72}
\bibitem{Laplaza72}
\bibinfo{author}{Miguel~L. \surnamestart Laplaza\surnameend}
  (\bibinfo{year}{1972}): \emph{\bibinfo{title}{Coherence for distributivity}}.
\newblock In \bibinfo{editor}{Saunders \surnamestart Mac~Lane\surnameend},
  editor: {\slshape \bibinfo{booktitle}{Coherence in Categories}}, {\slshape
  \bibinfo{series}{Lecture Notes in Mathematics}} \bibinfo{volume}{281},
  \bibinfo{publisher}{Springer-Verlag}, pp. \bibinfo{pages}{29--65},
  \doi{10.1007/BFb0059555}.

\bibitemdeclare{article}{Oreshkov2012}
\bibitem{Oreshkov2012}
\bibinfo{author}{Ognyan \surnamestart Oreshkov\surnameend},
  \bibinfo{author}{Fabio \surnamestart Costa\surnameend} \&
  \bibinfo{author}{\v{C}aslav \surnamestart Brukner\surnameend}
  (\bibinfo{year}{2012}): \emph{\bibinfo{title}{Quantum correlations with no
  causal order}}.
\newblock {\slshape \bibinfo{journal}{Nature Communications}}
  \bibinfo{volume}{3}, p. \bibinfo{pages}{1092}, \doi{10.1038/ncomms2076}.
\newblock \eprint{1105.4464}.

\bibitemdeclare{article}{Selinger2007CPM}
\bibitem{Selinger2007CPM}
\bibinfo{author}{Peter \surnamestart Selinger\surnameend}
  (\bibinfo{year}{2007}): \emph{\bibinfo{title}{Dagger compact closed
  categories and completely positive maps}}.
\newblock {\slshape \bibinfo{journal}{Electronic Notes in Theoretical Computer
  Science}} \bibinfo{volume}{170}, pp. \bibinfo{pages}{139--163},
  \doi{10.1016/j.entcs.2006.12.018}.

\bibitemdeclare{inproceedings}{simmons2022higherorder}
\bibitem{simmons2022higherorder}
\bibinfo{author}{Will \surnamestart Simmons\surnameend} \&
  \bibinfo{author}{Aleks \surnamestart Kissinger\surnameend}
  (\bibinfo{year}{2022}): \emph{\bibinfo{title}{Higher-Order Causal Theories
  Are Models of {BV}-Logic}}.
\newblock In: {\slshape \bibinfo{booktitle}{47th International Symposium on
  Mathematical Foundations of Computer Science (MFCS 2022)}}, {\slshape
  \bibinfo{series}{Leibniz International Proceedings in Informatics (LIPIcs)}}
  \bibinfo{volume}{241}, \bibinfo{publisher}{Schloss Dagstuhl--Leibniz-Zentrum
  f{\"u}r Informatik}, pp. \bibinfo{pages}{80:1--80:14},
  \doi{10.4230/LIPIcs.MFCS.2022.80}.
\newblock \eprint{2205.11219}.

\bibitemdeclare{misc}{Wilson2026HigherOrderCircuits}
\bibitem{Wilson2026HigherOrderCircuits}
\bibinfo{author}{Matt \surnamestart Wilson\surnameend} (\bibinfo{year}{2026}):
  \emph{\bibinfo{title}{Higher-order circuits}}.
\newblock \eprint{2602.18701}.

\bibitemdeclare{article}{Wilson2021HOPT}
\bibitem{Wilson2021HOPT}
\bibinfo{author}{Matt \surnamestart Wilson\surnameend} \&
  \bibinfo{author}{Giulio \surnamestart Chiribella\surnameend}
  (\bibinfo{year}{2021}): \emph{\bibinfo{title}{Causality in Higher Order
  Process Theories}}.
\newblock {\slshape \bibinfo{journal}{Electronic Proceedings in Theoretical
  Computer Science}} \bibinfo{volume}{343}, pp. \bibinfo{pages}{265--300},
  \doi{10.4204/EPTCS.343.12}.
\newblock \eprint{2107.14581}.

\bibitemdeclare{misc}{Wilson2022Framework}
\bibitem{Wilson2022Framework}
\bibinfo{author}{Matt \surnamestart Wilson\surnameend} \&
  \bibinfo{author}{Giulio \surnamestart Chiribella\surnameend}
  (\bibinfo{year}{2022}): \emph{\bibinfo{title}{A Mathematical Framework for
  Transformations of Physical Processes}}.
\newblock \eprint{2204.04319}.

\bibitemdeclare{misc}{WilsonChiribella2026MonoidalSU}
\bibitem{WilsonChiribella2026MonoidalSU}
\bibinfo{author}{Matt \surnamestart Wilson\surnameend} \&
  \bibinfo{author}{Giulio \surnamestart Chiribella\surnameend}
  (\bibinfo{year}{2026}): \emph{\bibinfo{title}{Monoidal su-categories}}.
\newblock \eprint{2608.15885}.

\bibitemdeclare{article}{WilsonChiribellaKissingerLocality}
\bibitem{WilsonChiribellaKissingerLocality}
\bibinfo{author}{Matt \surnamestart Wilson\surnameend}, \bibinfo{author}{Giulio
  \surnamestart Chiribella\surnameend} \& \bibinfo{author}{Aleks \surnamestart
  Kissinger\surnameend} (\bibinfo{year}{2026}): \emph{\bibinfo{title}{Quantum
  Supermaps are Characterized by Locality}}.
\newblock {\slshape \bibinfo{journal}{Quantum}} \bibinfo{volume}{10}, p.
  \bibinfo{pages}{2013}, \doi{10.22331/q-2026-03-09-2013}.
\newblock \eprint{2205.09844}.

\end{thebibliography}
\end{document}